\documentclass[12pt]{article}
\usepackage[paper=a4paper,margin=1in]{geometry}
\usepackage{t1enc}      

\usepackage{amsmath}
\usepackage{amsfonts}
\usepackage{amssymb}
\usepackage{amsthm}
\usepackage{graphicx}
\usepackage{mathrsfs}  

\newcommand{\edth} {\mbox{\symbol{'360}}}

\newcommand{\uA}{\underline A \,}

\theoremstyle{plain}

\providecommand{\keywords}[1]
{\small	\textbf{\textit{Keywords:}} #1 }

\numberwithin{equation}{section}

\begin{document}
\bibliographystyle{unsrt}

\title{An odd feature of the `most classical' states of $SU(2)$ 
invariant quantum mechanical systems}
\author{L\'aszl\'o B. Szabados \\
Wigner Research Centre for Physics, \\
H-1525 Budapest 114, P. O. Box 49, EU \\
e-mail: lbszab@rmki.kfki.hu}

\maketitle

\begin{abstract}
Complex and spinorial techniques of general relativity are used to determine 
all the states of the $SU(2)$ invariant quantum mechanical systems in which 
the equality holds in the uncertainty relations for the components of the 
angular momentum vector operator in two given directions. The expectation 
values depend on a discrete quantum number and two parameters, one of them 
is the angle between the two angular momentum components and the other is 
the quotient of the two standard deviations. Allowing the angle between the 
two angular momentum components to be arbitrary, \emph{a new genuine quantum 
mechanical phenomenon emerges}: it is shown that although the standard 
deviations change continuously, one of the expectation values changes 
\emph{discontinuously} on this parameter space. Since physically neither of 
the angular momentum components is distinguished over the other, this 
discontinuity suggests that the genuine parameter space must be a 
\emph{double cover} of this classical one: it must be diffeomorphic to a 
\emph{Riemann surface} known in connection with the complex function 
$\sqrt{z}$. Moreover, the angle between the angular momentum components 
plays the role of the parameter of an interpolation between the continuous 
range of the expectation values in the special case of the orthogonal 
angular momentum components and the discrete point spectrum of one angular 
momentum component. The consequences in the \emph{simultaneous} measurements 
of these angular momentum components are also discussed briefly. 
\end{abstract}



\section{Introduction}
\label{sec-1}

If the algebra of basic observables of the quantum system is the Heisenberg 
algebra, then the so-called canonical coherent states, defined to be the 
eigenstates of the annihilation operator, can be characterized in various 
equivalent ways (see e.g. \cite{Kl,ZFG}). E.g. these are the smallest 
uncertainty states for the basic canonically conjugate observables, and, 
with the Hamiltonian of the harmonic oscillator, they do not spread out in 
space during their time evolution. Thus, these states are usually 
interpreted as the `most classical' states of the quantum system. These 
states have particular significance e.g. in the characterization of the 
resulting state after the most accurate \emph{simultaneous} measurement of 
\emph{two non-commuting} observables \cite{ArKe,SHe}; or in the 
characterization of the system in the classical limit. 

The notion of coherent states has already been introduced for systems whose 
algebra of basic observables is more complicated than the Heisenberg algebra, 
e.g. when it is $su(2)$ (see e.g. \cite{Rad}-\cite{WaSaPa}). In addition to 
applications e.g. in quantum industry (in particular, in quantum optics see 
e.g. \cite{Kl,ZFG}, and in quantum computation see e.g. \cite{NiChu}), 
these states could have particular significance in basic science, e.g. in 
the characterization of the classical limit of the spin systems \cite{Lieb}, 
or in the determination of the large spin limit of the spin network model 
of the quantum spacetime suggested by Penrose in \cite{Pe}. 

Our aim in the present paper is to determine \emph{all} the `most classical' 
states of the general $SU(2)$-invariant quantum mechanical systems (even 
with given intrinsic spin) in the sense that, in these states, \emph{the 
equality holds in the uncertainty relation for the components of the angular 
momentum vector operator in two arbitrary, given directions}. The analogous 
problem was solved by Aragone \emph{et al} in \cite{Aretal} in the framework 
of the spin coherent states of Radcliffe \cite{Rad} in the special case when 
the two directions were orthogonal to one another (and the resulting states 
were called the `intelligent states'). As far as we know, the general case 
has not been considered hitherto. As we will see, allowing the angle between 
the two angular momentum components to be arbitrary, \emph{a new quantum 
mechanical phenomenon emerges}: the parameter space parameterizing the 
expectation values must be diffeomorphic to the non-trivial Riemann surface 
appearing in connection with the complex function $\sqrt{z}$ in complex 
analysis, rather than the \emph{classical} parameter space which is a simply 
connected subset of the complex plane. 

In deriving our results, complex and spinorial techniques of general 
relativity are used. It is these techniques that made it possible to 
determine the spectrum and the eigenfunctions of the relevant 
non-self-adjoint operator in explicit, \emph{closed form} in a relatively 
straightforward (but technically a bit involved) way. It is these techniques 
that can be (and, in fact, have been) used to carry out the analogous 
investigations in the considerably more complicated Euclidean invariant 
systems \cite{Sz2}. To make the present paper self-contained, these ideas 
and techniques are summarized in Appendix \ref{sub-A.1}. For a more detailed 
discussion of them, see e.g. \cite{PR}-\cite{FrSz}. 

In section \ref{sec-2}, we determine the most classical states of the 
$SU(2)$-invariant systems and discuss its properties, namely we calculate 
the expectation values, the standard deviations and find the wave functions. 
There we raise the possibility that the proper parameter space for the 
expectation value is the Riemann surface above. Section \ref{sec-3} is 
devoted to a brief discussion of how could it be possible to decide 
experimentally, at least in principle, that the genuine parameter space is 
this Riemann surface. Section \ref{sec-4} is a more detailed summary of the 
results and our conclusions. Several technical details of the analyses in 
section \ref{sec-2} are presented in Appendix \ref{sub-A.2}. 

Our conventions are mostly those of \cite{PR,HT}, except that the signature 
of the metric of the Euclidean 3-space is positive definite. Here we do 
\emph{not} use abstract indices: every index is a concrete (name, or 
component) index referring to some basis. Round brackets around indices 
denote symmetrization.

\section{The most classical states}
\label{sec-2}

As is well known, using the Cauchy--Schwarz inequality, for the standard 
deviation (or uncertainty) $\Delta_\phi{\bf A}$ and $\Delta_\phi{\bf B}$ of 
any two observables ${\bf A}$ and ${\bf B}$ in any normalized state $\phi$, 
respectively, Schr\"odinger derived the inequality 
\begin{equation}
(\Delta_\phi{\bf A})^2(\Delta_\phi{\bf B})^2\geq\frac{1}{4}\vert\langle
[{\bf A},{\bf B}]\rangle_\phi\vert^2+\Bigl(\langle\frac{1}{2}({\bf A}{\bf B}
+{\bf B}{\bf A})\rangle_\phi-\langle{\bf A}\rangle_\phi\langle{\bf B}\rangle
_\phi\Bigr)^2. \label{eq:2.0.1}
\end{equation}
Here, e.g. $\langle{\bf A}\rangle_\phi$ is the expectation value of ${\bf A}$, 
and the standard deviation in the state $\phi$ is given by $\Delta_\phi{\bf 
A}=\sqrt{\langle{\bf A}^2\rangle_\phi-(\langle{\bf A}\rangle_\phi)^2}$. The 
necessary and sufficient condition of the equality in (\ref{eq:2.0.1}) is 
well known to be the vanishing of the second term on the right, which is the 
first order correlation between the two expectation values, and the equality 
in the Cauchy--Schwarz inequality. These two conditions together yield that 
the state $\phi$ must be a solution of the eigenvalue equation 
\begin{equation}
({\bf A}-{\rm i}\lambda{\bf B})\phi=(\langle{\bf A}\rangle_\phi-{\rm i}
\lambda\langle{\bf B}\rangle_\phi)\phi,  \label{eq:2.0.2}
\end{equation}
where, for \emph{non-commuting} observables, $\lambda$ is a nonzero real 
constant. Then, as (\ref{eq:2.0.2}) immediately implies, $\vert\lambda\vert=
\Delta_\phi{\bf A}/\Delta_\phi{\bf B}$ holds; and, for the sake of simplicity, 
$\lambda$ will be chosen to be \emph{positive}. 

In the present paper, we use these general ideas to find the most classical 
states with a given intrinsic spin when the basic observables form the 
$su(2)$ Lie algebra. In the companion paper \cite{Sz2}, we consider systems 
for which this is the Lie algebra $e(3)$ of the Euclidean group.

\subsection{$SU(2)$ invariant elementary systems}
\label{sub-2.1}

The generators of the Lie algebra $su(2)$ are denoted by ${\bf J}_i$, $i=
1,2,3$, satisfying $[{\bf J}_i,{\bf J}_j]={\rm i}\hbar\varepsilon_{ijk}{\bf J}
^k$, where the lowering and raising of the small Latin indices are defined 
by $\delta_{ij}$ and its inverse, and $\varepsilon_{ijk}$ is the alternating 
Levi-Civita symbol. Its only Casimir operator is ${\bf J}_i{\bf J}^i$. 

We search for the general most classical states in the form of 
superpositions of such states belonging to the  \emph{irreducible} 
representations of $SU(2)$ with given, but arbitrary intrinsic spin. The 
finite dimensional irreducible representations of $SU(2)$ and $su(2)$ in 
their traditional, purely algebraic bra-ket formalism by unitary and 
self-adjoint operators, respectively, are well known. Nevertheless, the 
representations that we use here are less well known outside the general 
relativity community, and they are based on the complex line bundles 
${\cal O}(-2s)$ over the unit 2-sphere ${\cal S}\approx S^2$, where $2s\in
\mathbb{Z}$ is fixed. The integral of the first Chern class of ${\cal O}
(-2s)$ is $2s$. This is a topological invariant, characterizing the global 
non-triviality (or `twist') of the bundle, and it represents the intrinsic 
spin of the system. The carrier space of such a representation is an 
appropriate \emph{finite} dimensional subspace of the Hilbert space 
${\cal H}_s=L_2({\cal S},{\rm d}{\cal S})$ of the square-integrable cross 
sections $\phi$ of the line bundle, where the measure ${\rm d}{\cal S}$ is 
the natural metric area element on ${\cal S}$. The reason of using this 
less well known form of the unitary, irreducible representations of $su(2)$ 
is twofold: first, the \emph{intrinsic spin} of the system appears naturally 
in this formalism (see Appendix \ref{sub-A.1.4}), although it is not linked 
to any Casimir operator of $su(2)$; and, second, it is this form that fits 
smoothly to the most convenient representations of the $E(3)$-invariant 
elementary quantum mechanical systems that we will consider in \cite{Sz2}, 
where the results of the present paper will also be used. 

The form of the angular momentum operators that we use are (densely defined) 
self-adjoint differential operators acting on the smooth cross sections of 
${\cal O}(-2s)$. Explicitly, they are 
\begin{equation}
{\bf J}^i\phi=s\hbar n^i\phi+\hbar\bigl(m^i\,{\edth}'\phi-\bar m^i\,{\edth}
\phi\bigr), 
\label{eq:2.1.1}
\end{equation}
where $n^i$ are the Cartesian components of the unit normal of ${\cal S}$ in 
$\mathbb{R}^3$, $m^i$ and $\bar m^i$ are those of the complex null tangents 
of ${\cal S}$, and ${\edth}$ and ${\edth}'$, the so-called edth operators, 
are the covariant directional derivative operators in the corresponding 
complex null directions. The spin weight $s$ of $\phi$, i.e. the intrinsic 
spin of the system, will be fixed in what follows, and it may take any 
integer or half-odd-integer value. 

In this representation, the $su(2)$ Casimir operator is ${\bf J}_i{\bf J}^i
\phi=s^2\hbar^2\phi-\hbar^2({\edth}{\edth}'+{\edth}'{\edth})\phi$, where 
$({\edth}{\edth}'+{\edth}'{\edth})$ is just the familiar Laplacian on the 
unit 2-sphere. In Appendix \ref{sub-A.1}, we summarize the construction and 
the key properties of the null tangents, the line bundles ${\cal O}(-2s)$, 
the operators ${\edth}$ and ${\edth}'$ and the related concepts, as well as 
the derivation of (\ref{eq:2.1.1}) and the form of the Casimir operator 
(see, in particular, equations (\ref{eq:A.1.10})-(\ref{eq:A.1.11})). A more 
detailed discussion of the line bundles ${\cal O}(-2s)$ and the related 
concepts are given e.g. in \cite{EaTod} (see also \cite{PR,HT}). Appendix 
\ref{sub-A.1.1} and \ref{sub-A.1.5} are also intended to provide some 
informal `dictionary' to translate the geometrical notion used here to the 
more familiar ones in quantum mechanics/quantum technology.

\subsection{The spectrum}
\label{sub-2.2}

For any $\alpha^i,\beta^i\in\mathbb{R}^3$ satisfying $\alpha^i\alpha^j\delta
_{ij}=\beta^i\beta^j\delta_{ij}=1$, we form the operators ${\bf J}(\alpha):=
\alpha^i{\bf J}_i$ and ${\bf J}(\beta):=\beta^i{\bf J}_i$, i.e. the components 
of the angular momentum vector operator determined by the directions 
$\alpha^i$ and $\beta^i$. However, without loss of generality, we may assume 
that e.g. $\beta^1=\beta^2=0$ and $\beta^3=1$, because by an appropriate 
rotation of the Cartesian coordinate system this can always be achieved. 
Moreover, we assume that $\alpha_3\not=\pm1$, because otherwise $\alpha^i=
\pm\beta^i$ would be allowed, and for these ${\bf J}(\alpha)=\pm{\bf J}
(\beta)$ would hold. Then by (\ref{eq:2.0.1}) and the commutation relations 
we have $\Delta_\phi{\bf J}(\alpha)\Delta_\phi{\bf J}(\beta)\geq(\hbar/2)
\vert\alpha^i\beta^j\varepsilon_{ijk}\langle{\bf J}^k\rangle_\phi\vert$, in 
which, by (\ref{eq:2.0.2}), the equality holds precisely when $\phi$ solves 
the eigenvalue equation 
\begin{equation}
\bigl({\bf J}(\alpha)-{\rm i}\lambda{\bf J}(\beta)\bigr)\phi=\Bigl(\langle
{\bf J}(\alpha)\rangle_\phi-{\rm i}\lambda\langle{\bf J}(\beta)\rangle_\phi
\Bigr)\phi=:\hbar C\phi \label{eq:2.2.1}
\end{equation}
for some $\lambda>0$. By (\ref{eq:2.1.1}) this eigenvalue equation is 
\begin{equation}
\bigl(\alpha^i-{\rm i}\lambda\beta^i\bigr)\Bigl(m_i{\edth}'\phi-\bar m_i
{\edth}\phi+sn_i\phi\Bigr)=C\phi. \label{eq:2.2.2}
\end{equation}
We solve this eigenvalue problem in two steps: first, using spinorial 
techniques of general relativity, we determine the spectrum of the operator 
${\bf J}(\alpha)-{\rm i}\lambda{\bf J}(\beta)$, and then, in the next 
subsection, we determine the eigenfunctions both as series and \emph{in a 
closed form}, too. 

Since ${\bf J}_i{\bf J}^i$ and ${\bf J}(\alpha)-{\rm i}\lambda{\bf J}(\beta)$ 
are commuting, they have a system of common eigenfunctions. Hence, it might 
appear to be natural to search for the eigenfunctions of the latter in the 
form of a linear combination of the spin weighted spherical harmonics ${}_s
Y_{j,m}$, where $j=\vert s\vert,\vert s\vert+1,\vert s\vert+2,...$ and $m=-j,
-j+1,...,j$ (see Appendix \ref{sub-A.1.1}-\ref{sub-A.1.3}). However, the 
direct use of these harmonics in (\ref{eq:2.2.2}) would yield a coupled 
system of algebraic equations even for given $j$, and it would be difficult 
to get its solutions in a simple, closed form. Our strategy, suggested by 
Paul Tod \cite{Tod.private}, is based on the trick that, instead of the 
Cartesian spinor basis (see Appendix \ref{sub-A.1.1}), we use the 
\emph{principal spinors} of the (unitary spinor form $\gamma_{AB}$ of the) 
purely spatial complex vector $\alpha_i-{\rm i}\lambda\beta_i$. This makes 
the calculation of the spectrum of ${\bf J}(\alpha)-{\rm i}\lambda{\bf J}
(\beta)$ easier, giving it in a \emph{simple closed form}. The next two 
paragraphs are mostly based on Paul Tod's ideas and calculations. 

Thus, we form $\gamma_{AA'}:=(\alpha_i-{\rm i}\lambda\beta_i)\sigma^i_{AA'}$, 
and, since $\alpha_i-{\rm i}\lambda\beta_i$ is a spatial vector, its 
unitary spinor form, $\gamma_{AB}:=\gamma_{AA'}\sqrt{2}\sigma^{0A'}{}_B$, is 
symmetric. (Here, $\sigma^a_{AA'}=(\sigma^0_{AA'},\sigma^i_{AA'})$ are the 
standard $SL(2,\mathbb{C})$ Pauli matrices, including the factor $1/
\sqrt{2}$, and raising and lowering of the capital Latin indices are 
defined by the symplectic metric $\varepsilon^{AB}$ and its inverse. For 
our conventions, see Appendix \ref{sub-A.1.1}.) Then (\ref{eq:2.2.2}) 
takes the form 
\begin{eqnarray}
C\phi\!\!\!\!&=\!\!\!\!&\gamma_{AA'}o^A\bar\iota^{A'}{\edth}'\phi-\gamma_{AA'}
  \iota^A\bar o^{A'}{\edth}\phi+\sqrt{2}s\gamma_{AA'}o^A\bar o^{A'}\phi
  \nonumber \\
\!\!\!\!&=\!\!\!\!&-\gamma_{AB}o^Ao^B{\edth}'\phi-\gamma_{AB}\iota^A\iota^B
  {\edth}\phi+\sqrt{2}s\gamma_{AB}o^A\iota^B\phi, \label{eq:2.2.3}
\end{eqnarray}
where $o^A$ and $\iota^A$ are the vectors of the Newman--Penrose spinor 
basis (see Appendix \ref{sub-A.1.1}-\ref{sub-A.1.2}). Next, for any given 
$j$, let us consider the function 
\begin{equation*}
\phi=\phi_{A_1...A_{2j}}o^{A_1}\cdots o^{A_{j+s}}\iota^{A_{j+s+1}}\cdots\iota^{A_{2j}}
\end{equation*}
defined on ${\cal S}$, where the coefficients $\phi_{A_1...A_{2j}}$ are constant 
and completely symmetric in its indices. Recalling how the operators 
${\edth}$ and ${\edth}'$ act on the spinors $o^A$ and $\iota^A$ (see 
Appendix \ref{sub-A.1.3}) and how the operator ${\bf J}_i{\bf J}^i$ is 
built from ${\edth}$ and ${\edth}'$ (see Appendix \ref{sub-A.1.4}), it is 
easy to check that these functions are, in fact, eigenfunctions of ${\bf J}
_i{\bf J}^i$ with eigenvalue $j(j+1)\hbar^2$. Substituting this $\phi$ into 
(\ref{eq:2.2.3}) and using that the symplectic metric $\varepsilon^{AB}$ 
on the space $\mathbb{S}_A$ of spinors can be written as $\varepsilon
^{AB}=o^A\iota^B-\iota^Ao^B$, after some algebra we obtain 
\begin{eqnarray}
\!\!\!\!&{}\!\!\!\!&C\phi_{A_1...A_{2j}}o^{A_1}\cdots o^{A_{j+s}}\iota^{A_{j+s+1}}
  \cdots\iota^{A_{2j}} \label{eq:2.2.4} \\
\!\!\!\!&{}\!\!\!\!&=-\frac{1}{\sqrt{2}}\Bigl((j+s)\gamma_{BA_1}\phi^B
  {}_{A_2...A_{2j}}+(j-s)\gamma_{BA_{2j}}\phi^B{}_{A_1...A_{2j-1}}\Bigr)o^{A_1}\cdots 
  o^{A_{j+s}}\iota^{A_{j+s+1}}\cdots\iota^{A_{2j}}. \nonumber 
\end{eqnarray}
Since $\gamma_{AB}$ is symmetric, there are spinors $\mu_A$ and $\nu_A$, 
the so-called principal spinors of $\gamma_{AB}$, such that $\gamma_{AB}=
\sqrt{2}\mu_{(A}\nu_{B)}$. Substituting this into (\ref{eq:2.2.4}) we find 
\begin{equation}
C\phi_{A_1...A_{2j}}=-j\Bigl(\mu_A\phi^A{}_{(A_1...A_{2j-1}}\nu_{A_{2j})}+\nu_A
\phi^A{}_{(A_1...A_{2j-1}}\mu_{A_{2j})}\Bigr). 
\label{eq:2.2.5}
\end{equation}
We solve this equation when $\gamma_{AB}$ is not null (`generic case'), 
i.e. if $\mu_A$ and $\nu_A$ are not proportional to each other, and 
when $\gamma_{AB}$ is null (`exceptional case') separately. 

If $\gamma_{AB}$ is not null, then $0\not=(\nu_A\mu^A)^2=-\gamma_{AB}\gamma
^{AB}=-\gamma_{AA'}\gamma^{AA'}=(\alpha_i-{\rm i}\lambda\beta_i)(\alpha^i-
{\rm i}\lambda\beta^i)=1-\lambda^2-2{\rm i}\lambda\alpha_3$, and $\mu_A$ 
and $\nu_A$ span the space $\mathbb{S}_A$. Hence the spinors of the form 
\begin{equation}
\phi_{A_1...A_{2j}}=\mu_{(A_1}\cdots\mu_{A_{j-m}}\nu_{A_{j-m+1}}\cdots\nu_{A_{2j})}, 
\label{eq:2.2.6}
\end{equation}
$m=-j,-j+1,...,j$, form a basis in the space $\mathbb{S}_{(A_1...A_{2j})}$ of 
the totally symmetric spinors of rank $2j$ (see also Appendix 
\ref{sub-A.1.1}). Substituting this into (\ref{eq:2.2.5}), we obtain 
\begin{equation*}
C\mu_{(A_1}\cdots\mu_{A_{j-m}}\nu_{A_{j-m+1}}\cdots\nu_{A_{2j})}=m(\nu_A\mu^A)\mu_{(A_1}
\cdots\mu_{A_{j-m}}\nu_{A_{j-m+1}}\cdots\nu_{A_{2j})};
\end{equation*}
i.e. the eigenvalues are 
\begin{equation}
C=m(\nu_A\mu^A)=\pm m\sqrt{1-\lambda^2-2{\rm i}\lambda\alpha_3}, 
\label{eq:2.2.7}
\end{equation}
and the corresponding (not normalized) eigenfunctions are 
\begin{equation}
\phi_{s,j,m}:=\mu_{(A_1}\cdots\mu_{A_{j-m}}\nu_{A_{j-m+1}}\cdots\nu_{A_{2j})}o^{A_1}
\cdots o^{A_{j+s}}\iota^{A_{j+s+1}}\cdots\iota^{A_{2j}}. \label{eq:2.2.8}
\end{equation}
For the sake of concreteness, in the rest of the paper, we will choose the 
\emph{upper sign} in (\ref{eq:2.2.7}). 

The principal spinors of $\gamma_{AB}$ are determined only up to the scale 
ambiguity, $(\mu_A,\nu_A)\mapsto(\chi\mu_A,\chi^{-1}\nu_A)$, where $\chi$ is 
any non-zero complex constant; and up to their order. However, $\nu_A\mu^A$ 
is invariant under the rescalings by $\chi$, while under the interchange 
of the principal spinors both $\nu_A\mu^A$ and $m$ change sign (see 
(\ref{eq:2.2.6})). Therefore, the eigenvalue $C$ is independent of these 
ambiguities (as it should be). Under such transformations the 
eigenfunctions change according to $\phi_{s,j,m}\mapsto\chi^{-2m}\phi_{s,j,m}$ 
and $\phi_{s,j,m}\mapsto\phi_{s,j,-m}$, respectively. The first of these 
ambiguities is reduced to a phase ambiguity by the normalization of the 
eigenfunctions. 

If $\gamma_{AB}=\sqrt{2}\mu_A\mu_B$ (i.e. when $\gamma_{AB}$ is null), then 
let us consider the spinor of the form 
\begin{equation*}
\phi_{A_1...A_{2j}}=\mu_{(A_1}\cdots\mu_{A_n}\chi_{A_{n+1}...A_{2j})},
\end{equation*}
where $\chi_{A_1...A_{2j-n}}$ is some totally symmetric spinor for which 
$\mu^{A_1}\chi_{A_1...A_{2j-n}}\not=0$, $n=0,1,...,2j$. Substituting this into 
(\ref{eq:2.2.5}) we obtain 
\begin{eqnarray*}
C\mu_{(A_1}\cdots\mu_{A_n}\chi_{A_{n+1}...A_{2j})}\!\!\!\!&=\!\!\!\!&(2j+1)\mu_{(A_1}
  \cdots\mu_{A_{n+1}}\chi_{A_{n+2}...A_{2j}A)}\mu^A \\
\!\!\!\!&=\!\!\!\!&(2j-n)\mu_{(A_1}\cdots\mu_{A_{n+1}}\chi_{A_{n+2}...A_{2j})A}\mu^A.
\end{eqnarray*}
Since, however, the number of the spinors $\mu_A$ with free index on the 
left hand side is $n$ while on the right hand side it is $n+1$, this 
equation can hold true precisely when $n=2j$; i.e. the eigenvalue is zero, 
$C=0$, the constant spinor $\phi_{A_1...A_{2j}}$ is null with $\mu_A$ as its 
$2j$-fold principal spinor, and the corresponding (not  normalized) 
eigenfunction is 
\begin{equation}
\phi_{s,j}:=\mu_{A_1}\cdots\mu_{A_{2j}}o^{A_1}\cdots o^{A_{j+s}}\iota^{A_{j+s+1}}
\cdots\iota^{A_{2j}}. \label{eq:2.2.9}
\end{equation}
Thus, the solution of (\ref{eq:2.2.5}) in the exceptional case is \emph{not} 
the special case of that in the generic case: while the eigenvalue is the 
$m=0$ special case of (\ref{eq:2.2.7}), the eigenfunction is the $m=-j$ 
special case of (\ref{eq:2.2.8}). 

Remarkably enough, the eigenvalue (\ref{eq:2.2.7}) does \emph{not} directly 
depend on $j$ or on $s$; it depends on them only indirectly through $m$ 
since $m=-j,-j+1,...,j$ and $j=\vert s\vert,\vert s\vert+1,...$. Moreover, 
we got \emph{no} restriction on the parameters $\alpha_3$ and $\lambda$. 
Therefore, even with fixed $s$, there are infinitely many eigenfunctions, 
labelled by $j$, with the given eigenvalue (\ref{eq:2.2.7}). Let $\phi$ be 
\emph{any} normalized solution of the eigenvalue equation (\ref{eq:2.2.1}) 
with the given eigenvalue. Then by (\ref{eq:2.2.7}) $\langle{\bf J}(\alpha)
\rangle_\phi-{\rm i}\lambda\langle{\bf J}(\beta)\rangle_\phi=m\hbar\sqrt{1-
\lambda^2-2{\rm i}\lambda\alpha_3}$ holds. Since both ${\bf J}(\alpha)$ and 
${\bf J}(\beta)$ are self-adjoint, their expectation value must be real, 
and hence an elementary algebraic calculation yields that 
\begin{eqnarray}
\langle{\bf J}(\alpha)\rangle_\phi\!\!\!\!&=\!\!\!\!&m\frac{\hbar}
  {\sqrt{2}}\sqrt{1-\lambda^2+\sqrt{(1-\lambda^2)^2+4\lambda^2\alpha^2_3}}, 
  \label{eq:2.2.10a}\\
\lambda\langle{\bf J}(\beta)\rangle_\phi\!\!\!\!&=\!\!\!\!&m\,{\rm sign}
  (\alpha_3)\frac{\hbar}{\sqrt{2}}\sqrt{\lambda^2-1+\sqrt{(1-\lambda^2)
  ^2+4\lambda^2\alpha^2_3}} \label{eq:2.2.10b}
\end{eqnarray}
if $\alpha_3\not=0$; these are $m\hbar\sqrt{1-\lambda^2}$ and 0, respectively, 
if $\alpha_3=0$ and $\lambda<1$; while these are 0 and $m\hbar\sqrt{\lambda^2
-1}$, respectively, if $\alpha_3=0$ and $\lambda>1$. In the exceptional case 
both these expectation values are vanishing. 

Therefore, apart from the discrete quantum number $m$, the expectation 
values are functions of two parameters: $\alpha_3\in(-1,1)$, as we expected, 
and $\lambda\in(0,\infty)$. The appearance of $\lambda$ in the 
\emph{expectation values} is a bit surprising, because, by its very meaning, 
it characterizes \emph{the ratio of the two standard deviations}. Now we 
discuss the behaviour of the expectation values as functions of these two 
parameters. 

First, for fixed non-zero $m$, $\langle{\bf J}(\alpha)\rangle_\phi$ and 
$\langle{\bf J}(\beta)\rangle_\phi$, as functions of $\lambda$, behave in 
complementary ways: $\vert\langle{\bf J}(\alpha)\rangle_\phi\vert=\vert
\langle{\bf J}(\beta)\rangle_\phi\vert$ precisely when $\lambda=1$; but in 
the $\lambda\to0$ limit $\langle{\bf J}(\alpha)\rangle_\phi\to m\hbar$ and 
$\langle{\bf J}(\beta)\rangle_\phi\to m\hbar\alpha_3$; while in the $\lambda
\to\infty$ limit $\langle{\bf J}(\alpha)\rangle_\phi\to m\hbar\vert\alpha_3
\vert$ and $\langle{\bf J}(\beta)\rangle_\phi\to{\rm sign}(\alpha_3)m\hbar$. 
The expectation values can be zero only if $\alpha_3=0$ (see above); and, 
for $\alpha_3\not=0$ and $m\not=0$, $\langle{\bf J}(\alpha)\rangle_\phi/m$, 
as a function of $\lambda$, is strictly monotonically \emph{decreasing} 
from $\hbar$ to $\hbar\vert\alpha_3\vert$, while ${\rm sign}(\alpha_3)
\langle{\bf J}(\beta)\rangle_\phi/m$ is strictly monotonically 
\emph{increasing} from $\hbar\vert\alpha_3\vert$ to $\hbar$. Both 
expectation values are smooth functions of $\lambda$. 

On the other hand, also for fixed non-zero $m$, although $\langle{\bf J}
(\alpha)\rangle_\phi$ is a continuous function of $\alpha_3$, but it is 
\emph{not differentiable} on the $\lambda>1$ portion of the $\alpha_3=0$ 
line (on which it is vanishing). $\langle{\bf J}(\beta)\rangle_\phi$ behaves 
just in the opposite way: it is \emph{not continuous} at $\alpha_3=0$ for 
$\lambda>1$, its left/right limit is $\mp m\hbar\sqrt{\lambda^2-1}/\lambda$; 
but its derivative with respect to $\alpha_3$ is zero there both from the 
$\alpha_3>0$ and the $\alpha_3<0$ directions. Thus, $\langle{\bf J}(\beta)
\rangle_\phi$ \emph{has a jump}, while its square, $(\langle{\bf J}(\beta)
\rangle_\phi)^2$, is differentiable there. Therefore, as functions on the 
parameter space ${\cal P}:=\{(\alpha_3,\lambda)\vert \alpha_3\in(-1,1),
\lambda\in(0,\infty)\}$, \emph{the two expectation values behave in 
asymmetric ways}. 

It follows from these properties that, for given $j$, the range of the 
expectation values $\langle{\bf J}(\alpha)\rangle_\phi$ and $\langle{\bf J}
(\beta)\rangle_\phi$ is a connected interval of $\mathbb{R}$, parameterized 
by $\lambda$, precisely when the directions $\alpha^i$ and $\beta^i$ are 
orthogonal to each other (see above). This case was considered by Aragone 
\emph{et al} in \cite{Aretal}. If $j$ may take any allowed value, then the 
range of the expectation values is the whole $\mathbb{R}$. If the two 
directions are not orthogonal to each other, then the range of them is a 
\emph{union of intervals} of length $\vert m\vert\hbar(1-\vert\alpha_3
\vert)$. In the limit $\vert\alpha_3\vert\to1$ (i.e. when ${\bf J}(\alpha)
\to\pm{\bf J}(\beta)$), these intervals are getting to be disjoint and, 
ultimately, shrink to points. Thus, in this limit, the range of the 
expectation values reduces to the discrete (point) spectrum of the operator 
${\bf J}(\beta)$. Therefore, $\alpha_3$ provides the parameter in an 
\emph{interpolation} between the purely continuous, connected range of the 
expectation values and the completely discrete spectrum of ${\bf J}(\beta)$. 

We return to the discussion of the expectation values and the structure of 
the proper parameter space in subsection \ref{sub-2.5} and section 
\ref{sec-3}.

\subsection{The eigenfunctions}
\label{sub-2.3}

The eigenfunctions $\phi_{s,j,m}$ and $\phi_{s,j}$ given by (\ref{eq:2.2.8}) 
and (\ref{eq:2.2.9}) in the generic and exceptional cases, respectively, 
are not normalized. The corresponding normalized eigenfunctions will be 
denoted by $W_{s,j,m}:=N_{s,j,m}\,\phi_{s,j,m}$ and $W_{s,j}:=N_{s,j}\,\phi_{s,j}$, 
respectively, where the factors of normalization, $N_{s,j,m}$ and $N_{s,j}$, 
(as well as the explicit coordinate form of the eigenfunctions) will be 
determined in Appendix \ref{sub-A.2.1} and \ref{sub-A.2.2}. However, as we 
will see there, although the functions $W_{s,j,m}$ and $W_{s',j',m'}$ are 
$L_2$-orthogonal to one another if $s$ and $s'$ or $j$ and $j'$ are different, 
$W_{s,j,m}$ and $W_{s,j,m'}$ are, in general, not. In fact, since (in 
contrast to ${\bf J}_3$ or ${\bf J}_i{\bf J}^i$) the operator ${\bf J}
(\alpha)-{\rm i}\lambda{\bf J}(\beta)$ is \emph{not} self-adjoint (and, in 
fact, not even normal), the orthogonality of its eigenfunctions with 
different eigenvalues does not follow. 

Since the eigenfunctions $W_{s,j,m}$ and $W_{s,j}$ belong to the eigenvalue 
labeled by $m$ and zero, respectively, the general solution $\phi$ of the 
eigenvalue problem (\ref{eq:2.2.2}) with given $s$ can be written as a 
combination of these functions: 
\begin{equation}
\phi=\sum_{j={\rm max}\{\vert s\vert,\vert m\vert\}}^\infty c^jW_{s,j,m}, \hskip 30pt 
{\rm or} \hskip 30pt \sum_{j=\vert s\vert}^\infty c^jW_{s,j}, \label{eq:2.3.1}
\end{equation}
respectively, where $\sum_j\vert c^j\vert^2=1$. Next we determine the 
eigenfunctions of the eigenvalue equation (\ref{eq:2.2.2}) \emph{in closed 
form}. 

Since, as we will see, all the functions in (\ref{eq:2.2.2}) are polynomial 
in the complex stereographic coordinates, and polynomial equations are easier 
to solve than trigonometrical ones, we search for its solutions in these 
coordinates. In addition, since $\alpha_i\alpha^i=1$, we can (and will) write 
$(\alpha_1,\alpha_2)=\sqrt{1-\alpha^2_3}(\cos\alpha,\sin\alpha)$, where 
$\alpha\in[0,2\pi)$. 

Using $\beta^i=(0,0,1)$ and the explicit form of $m^i$ and the edth operators 
(see Appendix \ref{sub-A.1.2} and \ref{sub-A.1.3}), equation (\ref{eq:2.2.2}) 
in the coordinate system $(\zeta,\bar\zeta)$ takes the form 
\begin{eqnarray}
\!\!\!\!&{}\!\!\!\!&\Bigl(\frac{1}{2}\sqrt{1-\alpha_3^2}\bigl(1-\exp[-2
  {\rm i}\alpha]\zeta^2\bigr)+(\alpha_3-{\rm i}\lambda)\exp[-{\rm i}\alpha]
  \zeta\Bigr)\exp[{\rm i}\alpha]\frac{\partial\ln\phi}{\partial\zeta} 
  \nonumber \\
-\!\!\!\!&{}\!\!\!\!&\Bigl(\frac{1}{2}\sqrt{1-\alpha_3^2}\bigl(1-\exp[2
  {\rm i}\alpha]\bar\zeta^2\bigr)+(\alpha_3-{\rm i}\lambda)\exp[{\rm i}
  \alpha]\bar\zeta\Bigr)\exp[-{\rm i}\alpha]\frac{\partial\ln\phi}{\partial
  \bar\zeta} \nonumber \\
+\!\!\!\!&{}\!\!\!\!&\frac{1}{2}s\Bigl(\sqrt{1-\alpha_3^2}\bigl(\exp[{\rm i}
  \alpha]\bar\zeta+\exp[-{\rm i}\alpha]\zeta\bigr)-2(\alpha_3-{\rm i}\lambda)
  \Bigr)=C. \nonumber
\end{eqnarray}
Thus, it seems useful to rotate the complex coordinates according to $\zeta
\mapsto\xi:=\exp[-{\rm i}\alpha]\zeta$, and in the new coordinates $(\xi,
\bar\xi)$ this equation simplifies to 
\begin{eqnarray}
\!\!\!\!&{}\!\!\!\!&\Bigl(\frac{1}{2}\sqrt{1-\alpha_3^2}(1-\xi^2)+(\alpha_3
 -{\rm i}\lambda)\xi\Bigr)\frac{\partial\ln\phi}{\partial\xi} 
 \nonumber \\
-\!\!\!\!&{}\!\!\!\!&\Bigl(\frac{1}
 {2}\sqrt{1-\alpha_3^2}(1-\bar\xi^2)+(\alpha_3-{\rm i}\lambda)\bar\xi\Bigr)
 \frac{\partial\ln\phi}{\partial\bar\xi}=C+s(\alpha_3-{\rm i}\lambda)-
 \frac{1}{2}s\sqrt{1-\alpha_3^2}(\xi+\bar\xi). \nonumber
\end{eqnarray}
Denoting the \emph{complex} vector field on the left of this equation by 
$X$, this equation can be written in the compact form 
\begin{equation}
X(\ln\phi)=C-\frac{1}{2}s\Bigl(\sqrt{1-\alpha^2_3}\xi-\alpha_3+{\rm i}\lambda
\Bigr)-\frac{1}{2}s\Bigl(\sqrt{1-\alpha^2_3}\bar\xi-\alpha_3+{\rm i}\lambda
\Bigr), \label{eq:2.3.2}
\end{equation}
i.e. the directional derivative of $\ln\phi$ in the direction $X$ is a given 
function on ${\cal S}$. Rewriting $X$ in the polar coordinate system, it 
becomes clear that its real part is $\lambda$-times the rotation generator 
about $\beta^i$, while its imaginary part is the rotation generator about 
$\alpha^i$. Since $\alpha^i\not=\pm\beta^i$, $X$ does not have any zero on 
${\cal S}$.

To find its solution, let us observe that 
\begin{eqnarray*}
-\bigl(\sqrt{1-\alpha^2_3}\xi-\alpha_3+{\rm i}\lambda\bigr)\!\!\!\!&=\!\!\!\!&
 \frac{\partial}{\partial\xi}\Bigl(\frac{1}{2}\sqrt{1-\alpha^2_3}(1-\xi^2)
 +(\alpha_3-{\rm i}\lambda)\xi\Bigr)\\
\!\!\!\!&=\!\!\!\!&X\Bigl(\ln\bigl(\frac{1}{2}\sqrt{1-\alpha^2_3}(1-\xi^2)+
(\alpha_3-{\rm i}\lambda)\xi\bigr)\Bigr),
\end{eqnarray*}
and hence the last two terms on the right of (\ref{eq:2.3.2}) can be written 
in the form 
\begin{equation*}
X\Bigl(\ln\bigl(\frac{\sqrt{1-\alpha^2_3}(1-\xi^2)+2(\alpha_3-{\rm i}\lambda)
\xi}{\sqrt{1-\alpha^2_3}(1-\bar\xi^2)+2(\alpha_3-{\rm i}\lambda)\bar\xi}\bigr)
^{s/2}\Bigr).
\end{equation*}
Introducing the notation 
\begin{equation}
\xi_{\pm}:=\frac{\alpha_3-{\rm i}\lambda\pm\sqrt{1-\lambda^2-2{\rm i}\lambda
\alpha_3}}{\sqrt{1-\alpha^2_3}}, \label{eq:2.3.3}
\end{equation}
this can be rewritten into the remarkably simple form 
\begin{equation}
X\Bigl(\ln\bigl(\frac{\sqrt{1-\alpha^2_3}(1-\xi^2)+2(\alpha_3-{\rm i}\lambda)
\xi}{\sqrt{1-\alpha^2_3}(1-\bar\xi^2)+2(\alpha_3-{\rm i}\lambda)\bar\xi}\bigr)
^{s/2}\Bigr)=X\Bigl(\ln\bigl(\frac{(\xi-\xi_+)(\xi-\xi_-)}{(\bar\xi-\xi_+)
(\bar\xi-\xi_-)}\bigr)^{s/2}\Bigr); \label{eq:2.3.4}
\end{equation}
while the vector field $X$ is 
\begin{equation}
X=-\frac{1}{2}\sqrt{1-\alpha^2_3}\Bigl(\bigl(\xi-\xi_+\bigr)\bigl(\xi-\xi_-)
\frac{\partial}{\partial\xi}-\bigl(\bar\xi-\xi_+\bigr)\bigl(\bar\xi-\xi_-
\bigr)\frac{\partial}{\partial\bar\xi}\Bigr). 
\label{eq:2.3.5}
\end{equation}
It might be worth noting that $\xi_\pm$ is in a one-to-one correspondence 
with the expectation value given by (\ref{eq:2.2.10b}):  
\begin{equation}
\vert\xi_\pm\vert^2=\frac{m\hbar\pm\langle{\bf J}_3\rangle_\phi}{m\hbar\mp
\langle{\bf J}_3\rangle_\phi}, \label{eq:2.3.6}
\end{equation}
which can be inverted to express $\langle{\bf J}_3\rangle_\phi$ by $\vert\xi
_\pm\vert^2$. 

If $u$ were a function on ${\cal S}$ for which $X(u)=1$ held, then the first 
term on the right of (\ref{eq:2.3.2}) would have the form $X(Cu)$, and hence 
we already would have found a particular solution of the \emph{inhomogeneous 
equation} (\ref{eq:2.3.2}). We also need the general solution $\phi_0$ of 
the corresponding \emph{homogeneous} equation $X(\ln\phi_0)=0$, i.e. of $X
(\phi_0)=0$. To find this $u$, let us observe that if the functions $u_1=u_1
(\xi)$ and $u_2=u_2(\bar\xi)$ solve 
\begin{equation}
\frac{{\rm d}u_1}{{\rm d}\xi}=-\frac{1}{\sqrt{1-\alpha^2_3}}\frac{1}{(\xi-
\xi_+)(\xi-\xi_-)}, \hskip 20pt
\frac{{\rm d}u_2}{{\rm d}\bar\xi}=-\frac{1}{\sqrt{1-\alpha^2_3}}\frac{1}
{(\bar\xi-\xi_+)(\bar\xi-\xi_-)}, \label{eq:2.3.7}
\end{equation}
respectively, then $X(u_1-u_2)=1$ (and $X(u_1+u_2)=0$, too). However, since 
\begin{equation*}
\frac{\rm d}{{\rm d}\xi}\ln\bigl(\frac{\xi-\xi_+}{\xi-\xi_-}\bigr)=
\frac{\xi_+-\xi_-}{(\xi-\xi_+)(\xi-\xi_-)},
\end{equation*}
we find that 
\begin{equation}
u_1=-\frac{1}{2\sqrt{1-\lambda^2-2{\rm i}\lambda\alpha_3}}\ln\frac{\xi-\xi_+}
{\xi-\xi_-}, \hskip 20pt
u_2=-\frac{1}{2\sqrt{1-\lambda^2-2{\rm i}\lambda\alpha_3}}\ln\frac{\bar\xi-
\xi_+}{\bar\xi-\xi_-}, \label{eq:2.3.8}
\end{equation}
\emph{provided} that $1-\lambda^2-2{\rm i}\lambda\alpha_3\not=0$, i.e. when 
$(1-\lambda)^2+\alpha^2_3>0$ (`generic case'). Therefore, in the generic 
case, $u=u_1-u_2$ holds and the general \emph{local} solution of 
(\ref{eq:2.3.2}) is 
\begin{equation}
\phi=\phi_0\Bigl(\frac{(\xi-\xi_-)(\bar\xi-\xi_+)}{(\xi-\xi_+)(\bar\xi-\xi_-)}
\Bigr)^{m/2}\Bigl(\frac{(\xi-\xi_+)(\xi-\xi_-)}{(\bar\xi-\xi_+)(\bar\xi-\xi_-)}
\Bigr)^{s/2}, \label{eq:2.3.9}
\end{equation}
where we used the expression (\ref{eq:2.2.7}) (with the upper sign) for the 
eigenvalue $C$, and $\phi_0$ is the general solution of the homogeneous 
equation. Clearly, an arbitrary complex function $\phi_0$ of $v:=u_1+u_2$ 
solves $X(\phi_0)=0$; and it is easy to see that this is, in fact, the 
\emph{general} solution of $X(\phi_0)=0$. Indeed, let us define $\psi_0(u_1,
u_2):=\phi_0(\xi,\bar\xi)$ and observe that by (\ref{eq:2.3.5}) $X(\phi_0)
=0$ is equivalent to 
\begin{equation*}
0=\bigl(\xi-\xi_+\bigr)\bigl(\xi-\xi_-)\frac{\partial\phi_0}{\partial\xi}-
\bigl(\bar\xi-\xi_+\bigr)\bigl(\bar\xi-\xi_-\bigr)\frac{\partial\phi_0}
{\partial\bar\xi}=-\frac{1}{\sqrt{1-\alpha^2_3}}\Bigl(\frac{\partial\psi_0}
{\partial u_1}-\frac{\partial\psi_0}{\partial u_2}\Bigr). 
\end{equation*}
However, its general solution is an arbitrary complex valued smooth function 
of $v:=u_1+u_2$, i.e. $\phi_0$ is an arbitrary complex valued smooth function 
of 
\begin{equation*}
w:=\frac{(\xi-\xi_+)(\bar\xi-\xi_+)}{(\xi-\xi_-)(\bar\xi-\xi_-)}. 
\end{equation*}
In particular, 
\begin{equation*}
\omega:=\frac{(\xi-\xi_+)^a(\bar\xi-\xi_+)^a(\xi-\xi_-)^b(\bar\xi-\xi_-)^b}
{(1+\xi\bar\xi)^{a+b}}=w^a\Bigl(\frac{\xi_+-\xi_-}{\xi_+-\xi_- w}\Bigr)^{a+b}
\end{equation*}
is a solution of $X(\phi_0)=0$ even with \emph{arbitrary} real $a$ and $b$. 
However, with such a general $\phi_0$ (\ref{eq:2.3.9}) is only a \emph{local} 
solution of (\ref{eq:2.2.2}): we still have to ensure that $\phi$ be well 
defined even on small circles surrounding the poles, and be square 
integrable, too. These requirements restrict the structure of $\phi_0$, and, 
in particular, $a$ and $b$ in $\omega$ are restricted to be only non-negative 
integers or half-odd-integers. In particular, by (\ref{eq:A.2.3a}) and 
(\ref{eq:A.2.5a})-(\ref{eq:A.2.5b}) the eigenfunctions $W_{s,j,m}$ are, in 
fact, combinations of functions of the form (\ref{eq:2.3.9}) with $\phi_0=
\omega$ in which $a$ and $b$ are non-negative integer or half-odd-integer 
such that $a+b=j$. 

If $\lambda=1$ and $\alpha_3=0$ (`exceptional case'), then $\xi_\pm=-{\rm i}$ 
and the solution of equation (\ref{eq:2.3.7}) is given by 
\begin{equation}
u_1=\frac{1}{{\rm i}+\xi}, \hskip 20pt
u_2=\frac{1}{{\rm i}+\bar\xi}. \label{eq:2.3.10}
\end{equation}
Thus $u=u_1-u_2$, and the \emph{general} solution of $X(\phi_0)=0$ is an 
arbitrary smooth complex function of $v:=u_1+u_2$. Hence, in the exceptional 
case, the \emph{local} solution of (\ref{eq:2.3.2}) is 
\begin{equation}
\phi=\phi_0\bigl(\frac{{\rm i}+\xi}{{\rm i}+\bar\xi}\bigr)^s, 
\label{eq:2.3.11}
\end{equation}
where we have used that in the exceptional case the eigenvalue $C$ in 
(\ref{eq:2.2.7}) is zero. In particular, 
\begin{equation*}
\omega:=\Bigl(\frac{(\xi+{\rm i})(\bar\xi+{\rm i})}{(1+\xi\bar\xi)}\Bigr)^a
=\frac{1}{(1-{\rm i}v)^a}
\end{equation*}
is a solution of $X(\phi_0)=0$ even with \emph{arbitrary} real $a$. This 
$a$ is restricted to be a non-negative integer or half-odd-integer by the 
requirement that the corresponding $\phi$ be well defined and square 
integrable. In particular, by (\ref{eq:A.2.3b}) and (\ref{eq:A.2.5c}) the 
eigenfunctions $W_{s,j}$ are combinations of functions of the form 
(\ref{eq:2.3.11}) with $\phi_0=\omega$ in which $a=j$. 

Therefore, the eigenfunctions even with fixed eigenvalue depend on one 
(almost completely) free function; or, in their series expansion, on 
infinitely many freely specifiable complex constants $c^j$ (see 
(\ref{eq:2.3.1})).

\subsection{The standard deviations}
\label{sub-2.4}

Taking the $L_2$-scalar product of the eigenvalue equation ${\bf J}(\alpha)
\phi=\hbar C\phi+{\rm i}\lambda{\bf J}(\beta)\phi$ with itself and using the 
consequence $\langle\phi,{\bf J}(\alpha)\phi\rangle-{\rm i}\lambda\langle
\phi,{\bf J}(\beta)\phi\rangle=\hbar C$ of the eigenvalue equation, we 
obtain 
\begin{equation*}
\langle\phi,({\bf J}(\alpha))^2\phi\rangle=\langle{\bf J}(\alpha)\phi,
{\bf J}(\alpha)\phi\rangle=\langle\phi,{\bf J}(\alpha)\phi\rangle^2-\lambda^2
\langle\phi,{\bf J}(\beta)\phi\rangle^2+\lambda^2\langle\phi,({\bf J}
(\beta))^2\phi\rangle.
\end{equation*}
This does, in fact, imply that $\Delta_\phi{\bf J}(\alpha)=\lambda\Delta_\phi
{\bf J}(\beta)$, yielding $\Delta_\phi{\bf J}(\alpha)\Delta_\phi{\bf J}(\beta)
=\lambda(\Delta_\phi{\bf J}(\beta))^2$, too. Hence, to calculate the standard 
deviations, we need to compute only $\langle({\bf J}(\beta))^2\rangle_\phi$. 
Thus, let $\phi$ be given by (\ref{eq:2.3.1}). Then 
\begin{equation*}
\langle({\bf J}(\beta))^2\rangle_\phi=\langle{\bf J}(\beta)\phi,{\bf J}(\beta)
\phi\rangle=\sum_{j,j'={\rm max}\{\vert s\vert,\vert m\vert\}}^\infty\bar c^{j'}c^j
N_{s,j',m}N_{s,j,m}\langle{\bf J}_3\phi_{s,j',m},{\bf J}_3\phi_{s,j,m}\rangle.
\end{equation*}
Therefore, we should calculate $\langle{\bf J}_3\phi_{s,j',m},{\bf J}_3\phi
_{s,j,m}\rangle$. This, with the formal substitution $m=-j$, will give the 
general form of $\langle{\bf J}_3\phi_{s,j'},{\bf J}_3\phi_{s,j}\rangle$ in 
the exceptional case, too. 

To calculate $\langle{\bf J}_3\phi_{s,j',m},{\bf J}_3\phi_{s,j,m}\rangle$, we 
need to know the action of ${\bf J}_3$ on the eigenfunctions $\phi_{s,j,m}$. 
Repeating the analysis behind the equations 
(\ref{eq:2.2.4})-(\ref{eq:2.2.5}), we obtain that 
\begin{eqnarray}
{\bf J}_i\phi_{s,j,m}=\frac{\hbar}{\sqrt{2}}\Bigl\{\!\!\!\!&{}\!\!\!\!&
  (j+s)\sigma^B_i{}_A\phi_{BA_2...A_{2j}}o^Ao^{A_2}\cdots o^{A_{j+s}}\iota^{A_{j+s+1}}
  \cdots\iota^{A_{2j}} \nonumber \\
+\!\!\!\!&{}\!\!\!\!&(j-s)\sigma^B_i{}_A\phi_{BA_2...A_{2j}}\iota^Ao^{A_2}
  \cdots o^{A_{j+s+1}}\iota^{A_{j+s+2}}\cdots\iota^{A_{2j}}\Bigr\}. \label{eq:2.4.1}
\end{eqnarray}
As a by-product, this formula makes it possible to calculate all the 
expectation values $\langle{\bf J}_i\rangle_\phi$, too. In fact, using the 
techniques of Appendix \ref{sub-A.2.2}, a straightforward calculation yields 
\begin{equation}
\langle\phi_{s,j',m},{\bf J}_i\phi_{s,j,m}\rangle=-(-)^{2j}\delta_{j'j}\sqrt{2}
\hbar j4\pi\frac{(j-s)!(j+s)!}{(2j+1)!}\phi^\dagger_{BA_2...A_{2j}}
\sigma^B_i{}_C\phi^{CA_2...A_{2j}}, \label{eq:2.4.2}
\end{equation}
where the adjoint $\phi^\dagger_{A_1...A_{2j}}$ of the spinor $\phi_{A_1...A_{2j}}$ 
is defined according to the standard convention $\phi^\dagger_A:=\bar\phi_{A'}
\sqrt{2}\sigma^{A'}_{0A}$. Since this is non-zero only if $j'=j$, the 
expectation value of ${\bf J}_i$ in the state $\phi$ given by 
(\ref{eq:2.3.1}) is simply $\langle{\bf J}_i\rangle_\phi=\sum
_{j={\rm max}\{\vert s\vert,\vert m\vert\}}^\infty\vert c^j\vert^2N^2_{s,j,m}\langle
\phi_{s,j,m},{\bf J}_i\phi_{s,j,m}\rangle$. Using (\ref{eq:2.4.1}) and the 
techniques of Appendix \ref{sub-A.2.2}, the expectation values $\langle
\phi_{s,j,m},{\bf J}_i\phi_{s,j,m}\rangle$ can be expressed as 
\begin{eqnarray}
\langle\phi_{s,j,m},({\bf J}_1+{\rm i}{\bf J}_2)\phi_{s,j,m}\rangle\!\!\!\!&=
  \!\!\!\!&\hbar\,4\pi\frac{(j-s)!(j+s)!}{(2j+1)!}\sum_{k=0}^{2j}(2j-k)
  {{2j}\choose{k}}\Phi_{k,j,m}\bar\Phi_{k+1,j,m}, \nonumber \\
\langle\phi_{s,j,m},{\bf J}_3\phi_{s,j,m}\rangle\!\!\!\!&=\!\!\!\!&-\hbar\,4\pi 
  \frac{(j-s)!(j+s)!}{(2j+1)!}\sum_{k=0}^{2j}(j-k){{2j}\choose{k}}\vert
  \Phi_{k,j,m}\vert^2, \label{eq:2.4.3b}
\end{eqnarray}
where, analogously to (\ref{eq:2.2.8}) (see also Appendix \ref{sub-A.1.1}), 
we introduced the notation 
\begin{equation}
\Phi_{k,j,m}:=\phi_{A_1...A_{2j}}O^{A_1}\cdots O^{A_k}I^{A_{k+1}}\cdots I^{A_{2j}}.
\label{eq:2.4.4}
\end{equation}
These, together with the expression (\ref{eq:A.2.9}) for the factor of 
normalization and the explicit form (\ref{eq:A.2.10}) for $\Phi_{k,j,m}$ 
given in Appendix \ref{sub-A.2.2}, yield the general form for the expectation 
value of the operators ${\bf J}_i$ in the state $W_{s,j,m}$. In particular, 
the latter for ${\bf J}_3$ yields (\ref{eq:2.2.10b}) in that particular 
state. We will use this form of the expectation value of ${\bf J}_3$ in 
Appendix \ref{sub-A.2.4}. 

Using (\ref{eq:2.4.1}), a bit longer but straightforward spinorial 
calculation, similar to that behind (\ref{eq:2.4.2}), yields that $\langle
{\bf J}_3\phi_{s,j',m},{\bf J}_3\phi_{s,j,m}\rangle$ is zero unless $j'=j$, and 
that 
\begin{eqnarray*}
\langle{\bf J}_3\phi_{s,j,m},{\bf J}_3\phi_{s,j,m}\rangle=\!\!\!\!&{}
  \!\!\!\!&(-)^{2j}j\hbar^24\pi\frac{(j-s)!(j+s)!}{(2j+1)!}\Bigl(
  \beta^B{}_C\beta^C{}_D\phi^\dagger_{BA_2...A_{2j}}\phi^{DA_2...A_{2j}} \nonumber \\
\!\!\!\!&{}\!\!\!\!&+(2j-1)\beta^B{}_D\phi^\dagger_{BCA_3...A_{2j}}\beta^C{}_E
  \phi^{DEA_3...A_{2j}}\Bigr) \nonumber \\
=\!\!\!\!&{}\!\!\!\!&(-)^{2j}j\hbar^24\pi\frac{(j-s)!(j+s)!}{(2j+1)!}\Bigl(
  j\phi^\dagger_{A_1...A_{2j}}\phi^{A_1...A_{2j}}  \nonumber \\
\!\!\!\!&{}\!\!\!\!&+2(2j-1)O^AI^B\phi^\dagger_{ABA_3...A_{2j}}O_CI_D\phi
  ^{CDA_3...A_{2j}}\Bigr). 
\end{eqnarray*}
Here, in the second step, we used $I^AO_B-O^AI_B=\delta^A_B$ and $2\beta
^{AC}\beta_{CB}=-\delta^A_B$, where $\beta_{AB}$ is the unitary spinor form 
of $\beta_i$. Then, using the definition of the adjoint $\phi^\dagger
_{A_1...A_{2j}}$, we obtain 
\begin{eqnarray}
&{}&\hskip -20pt \langle{\bf J}_3\phi_{s,j,m},{\bf J}_3\phi_{s,j,m}
  \rangle \nonumber \\ 
&{}&=\hbar^2\,4\pi\frac{(j-s)!(j+s)!}{(2j+1)!}\,j^2\sum_{k=0}^{2j}{{2j}
  \choose{k}}\vert\phi_{{\uA}_1...{\uA}_{2j}}O^{A_1}\cdots O^{A_k}
  I^{A_{k+1}}\cdots I^{A_{2j}}\vert^2 \nonumber \\
&{}&-\hbar^2\,4\pi\frac{(j-s)!(j+s)!}{(2j+1)!}2j(2j-1)\sum_{k=0}^{2j-2}
  {{2j-2}\choose{k}} \nonumber \\
&{}&\hskip 25pt \times\vert O^AI^B\phi_{ABA_1...A_{2j-2}}O^{A_1}\cdots O^{A_k}
  I^{A_{k+1}}\cdots I^{A_{2j-2}}\vert^2 \nonumber \\
&{}&=\hbar^2\,4\pi\frac{(j-s)!(j+s)!}{(2j+1)!}\sum_{k=0}^{2j}(j-k)^2
  {{2j}\choose{k}}\vert\Phi_{k,j,m}\vert^2. \label{eq:2.4.5}
\end{eqnarray}
Since for any normalized state $\Delta_\phi{\bf J}(\alpha)\Delta_\phi{\bf J}
(\beta)=\lambda(\Delta_\phi{\bf J}(\beta))^2$ holds, the normalization 
condition $\sum_{j={\rm max}\{\vert s\vert,\vert m\vert\}}^\infty \vert c^j\vert^2=1$ 
and equation (\ref{eq:2.2.10b}) give that the `product uncertainty' in the 
state $\phi$ is 
\begin{equation}
\Delta_\phi{\bf J}(\alpha)\Delta_\phi{\bf J}(\beta)=\sum
_{j={\rm max}\{\vert s\vert,\vert m\vert\}}^\infty \vert c^j\vert^2\lambda\Bigl(
\langle{\bf J}_3W_{s,j,m},{\bf J}_3W_{s,j,m}\rangle-\langle W_{s,j,m},{\bf J}_3
W_{s,j,m}\rangle^2\Bigr). \label{eq:2.4.6}
\end{equation}
Hence, it is enough to evaluate the product uncertainty only in the 
normalized eigenstates $W_{s,j,m}$; i.e., by (\ref{eq:2.4.5}), to clarify 
the properties of $\Phi_{k,j,m}$ only. This will be done in Appendix 
\ref{sub-A.2.2}-\ref{sub-A.2.3}. In particular, in Appendix \ref{sub-A.2.3} 
we show that $\langle{\bf J}_3W_{s,j,m},{\bf J}_3W_{s,j,m}\rangle$ does not 
depend on the sign of $m$, while $\langle W_{s,j,m},{\bf J}_3W_{s,j,m}\rangle$ 
changes sign if $m$ is replaced by $-m$. Moreover, $\langle{\bf J}_3W_{s,j,m},
{\bf J}_3W_{s,j,m}\rangle$ is continuous even at $\alpha_3=0$ for $\lambda>1$, 
while $\langle W_{s,j,m},{\bf J}_3W_{s,j,m}\rangle$ has a jump there. In addition, 
in Appendix \ref{sub-A.2.4}, we determine the asymptotic behaviour of the 
standard deviations in the $\lambda\to0$ and $\lambda\to\infty$ limits. We 
will see that the standard deviation for ${\bf J}(\beta)$ is finite and 
that for ${\bf J}(\alpha)$ tends to zero as $\lambda$ if $\lambda\to0$; 
while the standard deviation for ${\bf J}(\beta)$ tends to zero as 
$1/\lambda$ and that for ${\bf J}(\alpha)$ is finite if $\lambda\to\infty$. 
Hence, the product uncertainty tends to zero in both limits. Therefore, the 
uncertainties do \emph{not} indicate any asymmetry between the ${\bf J}
(\alpha)$ and ${\bf J}(\beta)$ angular momentum components, just as we 
expect it on physical grounds. 

Finally, in the exceptional case, (\ref{eq:2.4.5}) together with the 
expression of $\mu_AO^A$ and $\mu_AI^A$ given in Appendix \ref{sub-A.2.2} 
yield $\langle{\bf J}_3W_{s,j},{\bf J}_3W_{s,j}\rangle=j\hbar^2/2$. Since 
$\langle W_{s,j},{\bf J}_3W_{s,j}\rangle$ $=0$, we find that the product 
uncertainty in the state $W_{s,j}$ is $j\hbar^2/2$. Hence, in the general 
eigenstate $\phi=\sum_{j=\vert s\vert}^\infty c^jW_{s,j}$, the product 
uncertainty is $(\hbar^2/2)\sum_{j=\vert s\vert}^\infty j\vert c^j\vert^2$, which 
can be zero only for $j=\vert s\vert=0$.

\subsection{The extension of the classical parameter space}
\label{sub-2.5}

Since physically ${\bf J}(\alpha)$ and ${\bf J}(\beta)$ are on equal footing, 
both should have the same qualitative properties. Indeed, the behaviour 
of the standard deviations does not break this symmetry between the two 
observables. However, the behaviour of the expectation values $\langle{\bf J}
(\alpha)\rangle_\phi$ and $\langle{\bf J}(\beta)\rangle_\phi$ as functions on 
the \emph{classical parameter space} ${\cal P}:=\{(\alpha_3,\lambda)\vert
\alpha_3\in(-1,1),\lambda\in(0,\infty)\}$ apparently \emph{does} break this 
symmetry. Hence, this odd behaviour indicates that either the expectation 
values should be extended to be \emph{double valued} on ${\cal P}$, or they 
should in fact be functions on a \emph{non-trivial Riemann surface} ${\cal 
R}$ larger than ${\cal P}$, which must be homeomorphic to the one known in 
elementary complex analysis in connection with the function $\sqrt{z}$. 

In fact, this second possibility seems more natural, because, for fixed $m$ 
and apart form the factor $\hbar$, the complex eigenvalue of ${\bf J}(\alpha)
-{\rm i}\lambda{\bf J}(\beta)$ is the \emph{square root} of the complex 
function $1-\lambda^2-2{\rm i}\lambda\alpha_3$ on ${\cal P}$ (see equation 
(\ref{eq:2.2.7})). The corresponding Riemann surface ${\cal R}$ is obtained 
from ${\cal P}$ by cutting it along the $\lambda\geq1$ part of the $\alpha_3
=0$ axis, and identifying the resulting edges for $\lambda>1$ with the 
opposite edges of another copy of ${\cal P}$ that has been cut in the similar 
way. The branch point of the resulting Riemann surface is at $\lambda=1$, 
$\alpha_3=0$, which corresponds just to the exceptional case of subsections 
\ref{sub-2.2} and \ref{sub-2.3}. Then the two expectation values are extended 
from the first to the second copy of ${\cal P}$ in ${\cal R}$ just to be 
$-1$ times the ones on the original `classical' copy ${\cal P}$. With this 
extension the resulting expectation values behave in the same way, and, in 
particular, will be differentiable on the whole of ${\cal R}$.

\section{On the simultaneous measurement of ${\bf J}(\alpha)$ 
and ${\bf J}(\beta)$}
\label{sec-3}

In the pioneering work \cite{ArKe}, Arthurs and Kelly raised the possibility 
of the \emph{simultaneous} measurement of the conjugate, and hence \emph{not 
commuting}, observables in a Heisenberg system. Soon later, it was argued 
\cite{SHe} that \emph{every} measurement of the momentum or of the position 
is, in fact, such a simultaneous measurement, although while we measure one 
quantity precisely, we measure the conjugate one rather imprecisely. For 
example,  when we do a precise measurement of the momentum of a particle, 
then the particle must be in the measuring apparatus (or at least in the 
laboratory), so we do a rude position measurement as well; and when we measure 
the position of the particle accurately and the measuring apparatus is not 
destroyed in the measuring procedure, then we can be sure that the particle's 
momentum could not have arbitrarily large momentum, so we did a rude momentum 
measurement, too. This idea of the simultaneous measurement of the conjugate 
observables was discussed further e.g. in \cite{SHe,St,Ra}, clarifying, 
in particular, the precise relationship between the errors of the measurements 
and the standard deviations, deriving inequalities for the former, etc. 

The basis of the possibility of such simultaneous measurements is that the 
non-commuting observables are \emph{different} kinds of quantities: to 
measure them we need different devices, and during the simultaneous running 
of them one does not make impossible to measure the other. If, however, 
the two non-commuting observables are of the \emph{same} kind, e.g. the 
${\bf J}(\alpha)$ and ${\bf J}(\beta)$ components of angular momentum, then 
it does not seem to be possible to carry out such a measurement \emph{in a 
direct way}. For example, in the Stern--Gerlach apparatus we cannot have 
two different magnetic fields at the same time to measure the spin of the 
particle in the corresponding two different directions. 

Nevertheless, the phenomenon of the quantum entanglement makes it possible, 
at least in principle, to measure ${\bf J}(\alpha)$ and ${\bf J}(\beta)$ 
simultaneously \emph{in an indirect way}. In fact, it is known that even the 
coherent spin states can be entangled \cite{WaSaPa}, and one particle in 
its most classical state can be maximally entangled with a second one. Thus, 
preparing two such particles to be maximally entangled and then spatially 
separating them, we can measure ${\bf J}(\alpha)$ for one, and ${\bf J}
(\beta)$ for the second in two independent Stern--Gerlach apparatus. 

In this way, in principle, one can verify experimentally whether or not 
the proper parameter space preferred by Nature is the classical ${\cal 
P}$ or the Riemann surface ${\cal R}$. (The parameter $\alpha_3$ has the 
obvious meaning in the experimental apparatus, and $\lambda$ can also be 
controlled in an indirect way since it is the quotient of the two standard 
deviations.) Indeed, let us consider the closed path in the classical 
parameter space ${\cal P}$ consisting of the following four straight line 
segments: 
\begin{description}
\item[1.] the initial point is given by $\alpha_3=1/\sqrt{2}$ and $\lambda=2$ 
(i.e. the angle between the directions $\alpha^i$ and $\beta^i$ is $\pi/4$, 
and $\Delta_\phi{\bf J}(\alpha)$ is just twice of $\Delta_\phi{\bf J}(\beta)$),
and the end point is defined by $\alpha_3=1/\sqrt{2}$ and $\lambda=1/2$; 
\item[2.] the end point of the second segment is at $\alpha_3=-1/\sqrt{2}$ 
and $\lambda=1/2$ (i.e. we enlarge the angle between $\alpha^i$ and $\beta^i$ 
from $\pi/4$ to $3\pi/4$ while keeping $\lambda$ to be $1/2$); 
\item[3.] the end point of the third segment is at $\alpha_3=-1/\sqrt{2}$ 
and $\lambda=2$; and 
\item[4.] the fourth segment closes the path, returning to the initial point 
of the first segment. 
\end{description}
Next consider a 1-parameter family of the most classical states with fixed 
non-zero $m$ and parameterized by the parameter of this closed path, and 
measure the two expectation values. If we find that \emph{both} expectation 
values change continuously along the closed path from their initial value at 
the initial point of the path to their own negative at the end point of the 
path, then the genuine parameter space is the Riemann surface ${\cal R}$; 
but if \emph{only one} changes continuously along the path but the other had 
a discontinuity (or, equivalently, one does \emph{not} change sign along the 
closed path but the other does), then the Nature's parameter space is the 
classical ${\cal P}$.

\section{Results, discussion and conclusions}
\label{sec-4}

In \cite{Aretal}, Aragone \emph{et al} determined the most classical 
states with respect to two Cartesian components of the angular momentum 
vector operator in quantum systems with the $su(2)$ as the algebra of its 
basic observables. These states with given total angular momentum quantum 
number $j$ and spin $s$ depend on a discrete quantum number $m$ and a 
single parameter $\lambda$. However, as the functions of $\lambda$, the 
expectation values of the two components change in an \emph{asymmetric} 
way: for $\lambda<1$ one is zero and the other is decreasing with $\lambda$, 
while for $\lambda>1$ the former is increasing with $\lambda$ while the 
latter is zero (see the discussion following equation (\ref{eq:2.2.10b})). 

In the present paper, we extended (and, in fact, completed) the 
investigations above: we determined \emph{all} the most classical states 
for the components of the angular momentum vector operator in \emph{any} 
two given directions. Allowing the angle between the two angular momentum 
components to be \emph{arbitrary}, we found that the expectation values and 
the standard deviations depend on one discrete quantum number, $m$, and two 
continuous parameters. In addition, the angle between the two directions 
provided an \emph{interpolation} between the continuous range of the 
expectation values found by Aragone \emph{et al} and the discrete point 
spectrum of one angular momentum component. 

Since the two expectation values are the real and imaginary parts of the 
square root of a complex function on this two-dimensional parameter space, 
one of the expectation values changes continuously but the other 
\emph{discontinuously} on this (classical) parameter space. As far as we 
know, this is a \emph{new quantum mechanical phenomenon}. The asymmetric 
behaviour of the expectation values in the special case considered by 
Aragone \emph{et al} and mentioned above is a special manifestation of this 
phenomenon. However, physically, neither of the angular momentum components 
is distinguished over the other; and the standard deviations do, in fact, 
show this symmetry. As a possible resolution of the contradiction between 
the symmetry of the angular momentum components and the apparently 
asymmetric behaviour of the \emph{expectation values} we suggest that 
\emph{the genuine parameter space should be diffeomorphic to the Riemann 
surface appearing in connection with the function $\sqrt{z}$ in complex 
analysis}. With this extension of the classical parameter space the 
asymmetric behaviour of the expectation values disappears. The odd 
behaviour of the expectation values on the classical parameter space seems 
to be analogous to that of the scissors in Dirac's scissors problem (see 
e.g. \cite{PR}, page 43); but, being independent of the value of the 
intrinsic spin, this is independent of the fermionic nature of the system. 
Based on the use of the entangled coherent states, we raise the possibility 
of a potential experimental verification of this phenomenon. 

\section{Acknowledgments}

Thanks are due to Paul Tod for the idea how the spectrum of the operator 
${\bf J}(\alpha)-{\rm i}\lambda{\bf J}(\beta)$ of subsection \ref{sub-2.2} 
can be calculated in the most economical way, as well as for his helpful 
remarks and comments; to P\'eter Vecserny\'es for his remarks on the 
classical limit of non-relativistic quantum systems; to Zolt\'an Zimbor\'as 
for the discussion of the quantum information theoretic interpretation of the 
various spinorial notions, as well as for suggesting reference \cite{NiChu}; 
and to the Referee for suggesting to add the `informal dictionary' in 
Appendix \ref{sub-A.1.5}. 

This research did not receive any specific grant from funding agencies in 
the public, commercial, or not-for-profit sectors. The author has no 
conflicts to disclose. 

\appendix
\section{Appendices}
\label{sec-A}

\subsection{The irreducible representations of $su(2)$ by spin 
weighted functions}
\label{sub-A.1}

In Appendices \ref{sub-A.1.1}--\ref{sub-A.1.4}, mostly to fix the notations 
and the technical background, we summarize the spinorial and complex techniques 
that we used in the calculations in the main part (as well as in Appendix 
\ref{sub-A.2}) of the paper. To make these spinorial notions more 
understandable for a wider readership, we also add Appendix \ref{sub-A.1.5} 
in which we show how some of these basic notions correspond to those e.g. 
in quantum optics and quantum information theory. (Thanks are due to one of 
the Referees for suggesting to provide such an informal `dictionary'.)

\subsubsection{An algebraic introduction of the spin weighted spherical 
harmonics}
\label{sub-A.1.1}

As is well known, any unitary irreducible (and hence finite dimensional) 
representation of $SU(2)$ can be labeled by a non-negative integer or 
half-odd-integer $j$, i.e. for which $2j=0,1,2,...$. The carrier space 
${\cal H}_j$ of such a representation is a $2j+1$ dimensional Hilbert space, 
in which the vectors of an orthonormal basis are usually denoted by $\vert j,
m\rangle$, where $m=-j,-j+1,...,j$. 

One concrete realization of this abstract Hilbert space can be based on the 
use of completely symmetric spinors. The space $\mathbb{S}^A$ of 
(two-component, unitary) spinors is a two-complex dimensional vector space 
endowed by a symplectic and a positive definite Hermitean metric that are 
compatible with each other. These metrics can be given in terms of the 
vectors $O_A$ and $I_A$ of a basis in $\mathbb{S}^A$: these, respectively, are 
$\varepsilon_{AB}:=O_AI_B-O_BI_A$ and $\sqrt{2}T_{AA'}:=O_A\bar O_{A'}+I_A\bar I
_{A'}$, where over-bar denotes complex conjugation. (For a general background, 
see e.g. \cite{PR,HT}).  Then, lowering and raising the spinor indices by 
$\varepsilon_{AB}$ and its inverse $\varepsilon^{AB}$, respectively, it is easy 
to check that this basis $\{O^A,I^A\}$ is normalized with respect to the 
symplectic and orthonormal with respect to the Hermitean metric: $O_AI^A=
\varepsilon^{AB}O_AI_B=1$ and $\sqrt{2}T_{AA'}O^A\bar O^{A'}=\sqrt{2}T_{AA'}I^A
\bar I^{A'}=1$, $\sqrt{2}T_{AA'}O^A\bar I^{A'}=0$. We call such a basis a 
Cartesian spin frame; which is unique only up to $SU(2)$ transformations. 
These imply that $T_{AA'}T^{AA'}=1$, $T_{AA'}\bar O^{A'}=-I_A/\sqrt{2}$ and 
$T_{AA'}\bar I^{A'}=O_A/\sqrt{2}$ also hold. With the choice $O^A=\delta^A_0$, 
$I^A=\delta^A_1$ this Hermitean metric is just the unit matrix $\sqrt{2}
\sigma^0_{AA'}$, where $\sigma^0_{AA'}$ denotes the zeroth of the four $SL(2,
\mathbb{C})$ Pauli matrices $\sigma^a_{AA'}$ (including the factor $1/
\sqrt{2}$), according to the conventions of \cite{PR}. With the 
identification $O^A=\vert\frac{1}{2},-\frac{1}{2}\rangle$, $I^A=\vert
\frac{1}{2},\frac{1}{2}\rangle$ the space $\mathbb{S}^A$ can be identified 
with the Hilbert space ${\cal H}_j$ above with $j=\frac{1}{2}$, and the 
Hermitean scalar product of the two states $\phi^A=\vert\phi\rangle$ and 
$\psi^A=\vert\psi\rangle$ is just $\langle\phi\vert\psi\rangle=\sqrt{2}
T_{AA'}\bar\phi^{A'}\psi^A$. Note, however, that while in quantum mechanics 
the relative phase of the vectors of an orthonormal basis is usually not 
fixed, the relative phase of $O^A$ and $I^A$ in $\mathbb{S}^A$ is fixed by 
the condition $O_AI^A=1$. 

The space $\mathbb{S}_{(A_1...A_{2j})}$ of the completely symmetric spinors 
$\phi_{A_1...A_{2j}}$ of rank $2j$, i.e. for which $\phi_{A_1...A_{2j}}=\phi
_{(A_1...A_{2j})}$, is the symmetrized $2j$-fold tensor product of the spin space 
$\mathbb{S}_A$ with itself. If $\{O_A,I_A\}$ is any basis in $\mathbb{S}_A$, 
then the spinors of the form $Z(j,m)_{A_1...A_{2j}}:=O_{(A_1}\cdots O_{A_{j-m}}$ 
$I_{A_{j-m+1}}\cdots I_{A_{2j})}$ form a basis in $\mathbb{S}_{(A_1...A_{2j})}$. If 
the basis $\{O_A,I_A\}$ is chosen as above, then the Hermitean scalar product 
makes it possible to normalize the basis vectors $Z(j,m)_{A_1...A_{2j}}$. In 
fact, the vectors 
\begin{equation*}
\sqrt{\frac{(2j)!}{(j+m)!(j-m)!}}Z(j,m)_{A_1...A_{2j}}
\end{equation*}
have not only unit norm with respect to $\sqrt{2}T^{AA'}$, but they are 
orthogonal to one another as well. Thus they can be identified with the 
abstract basis vectors $\vert j,m\rangle$ in ${\cal H}_j$ above. Hence, 
a completely symmetric spinor $\phi_{A_1...A_{2j}}\in\mathbb{S}_{(A_1...A_{2j})}$ 
can equivalently be represented by its components in this basis, or by the 
contractions $\phi_{A_1...A_{2j}}Z(j,m)^{A_1...A_{2j}}$, which are $2j+1$ complex 
numbers. 

Another basis in the space $\mathbb{S}_A$ is provided by the so-called 
Newman--Penrose spin frame $\{o_A,\iota_A\}$. Such a frame can be chosen to 
be 
\begin{equation}
o_A=\frac{-{\rm i}}{\sqrt{1+\zeta\bar\zeta}}\bigl(\zeta O_A+I_A\bigr),
\hskip 20pt
\iota_A=\frac{-{\rm i}}{\sqrt{1+\zeta\bar\zeta}}\bigl(O_A-\bar\zeta 
I_A\bigr), \label{eq:A.1.1}
\end{equation}
where $\zeta\in\mathbb{C}$ (see e.g. \cite{PR}). The matrix of the basis 
transformation $\{O_A,I_A\}\mapsto\{o_A,\iota_A\}$ is an $SU(2)$ matrix. 
Hence, in particular, $\sqrt{2}t_{AA'}:=o_A\bar o_{A'}+\iota_A\bar\iota_{A'}
=O_A\bar O_{A'}+I_A\bar I_{A'}=\sqrt{2}T_{AA'}$ holds. The frame $\{o_A,\iota
_A\}$ is a two-real-parameter family of bases, which can be considered to 
be defined on the (e.g. unit) 2-sphere ${\cal S}$, where $\zeta$ is the 
complex stereographic coordinate on the 2-sphere (see the next subsection). 
Hence, the spinors of the form $z(j,s)_{A_1...A_{2j}}:=o_{(A_1}\cdots o_{A_{j+s}}
\iota_{A_{j+s+1}}\cdots\iota_{A_{2j})}$, $s=-j,-j+1,...,j$, form another basis 
in $\mathbb{S}_{(A_1...A_{2j})}$. These basis vectors can also be normalized, 
and 
\begin{equation*}
\sqrt{\frac{(2j)!}{(j-s)!(j+s)!}}z(j,s)_{A_1...A_{2j}}
\end{equation*}
form another orthonormal basis with respect to $\sqrt{2}t^{AA'}=\sqrt{2}
T^{AA'}$. Hence, the contractions 
\begin{equation*}
U(j)_{m,s}:=\frac{(2j)!}{\sqrt{(j+m)!(j-m)!(j-s)!(j+s)!}}
Z(j,m)_{A_1...A_{2j}}z(j,s)^{A_1...A_{2j}}
\end{equation*}
form a $(2j+1)\times(2j+1)$ unitary matrix, taking one orthonormal basis 
of $\mathbb{S}_{(A_1...A_{2j})}$ to another one. Hence, the spinor $\phi
_{A_1...A_{2j}}\in\mathbb{S}_{(A_1...A_{2j})}$ can also be represented by the 
contractions $\phi_{A_1...A_{2j}}z(j,s)^{A_1...A_{2j}}$, which are $2j+1$ 
\emph{special complex valued functions on ${\cal S}$}. 

Apart from a numerical coefficient, the familiar spin weighted spherical 
harmonics are just the components of the matrix $U(j)_{m,s}$: 
\begin{equation}
{}_sY_{j,m}:=(-)^{j+m}\sqrt{\frac{2j+1}{4\pi}}U(j)_{m,s}, \label{eq:A.1.1Y}
\end{equation}
where the numerical coefficient makes these to be normalized with respect to 
the $L_2$-scalar product on the unit sphere; and, for $s=0$, these are just 
the familiar ordinary spherical harmonics $Y_{j,m}$. For further properties 
of these harmonics, see \cite{NP,Goetal,PR,HT}.

\subsubsection{Spinorial coordinates on ${\cal S}$}
\label{sub-A.1.2}

Let $p^i$, $i=1,2,3$, denote Cartesian coordinates in $\mathbb{R}^3$, in 
which the components of the independent rotation Killing vector fields are 
$k^i_{mn}=p^j(\delta_{jn}\delta^i_m-\delta_{jm}\delta^i_n)$, $m,n=1,2,3$. Thus 
here $i$ is the \emph{vector} index, while $m$ and $n$ are the \emph{name} 
indices of the Killing fields. These are tangent to the 2-spheres of radius 
$P$, ${\cal S}:=\{p^i\in\mathbb{R}^3\vert P^2:=\delta_{ij}p^ip^j={\rm const}\}$, 
and vanish at the point $p^i=0$. These Killing fields with the Lie bracket 
generate the Lie algebra $su(2)$. 

The complex stereographic coordinates, projected from the \emph{north} pole, 
are defined on $U_n:={\cal S}-\{(0,0,P)\}$, the sphere minus its \emph{north} 
pole, by $\zeta:=\exp({\rm i}\varphi)\cot(\theta/2)$, where $(\theta,
\varphi)$ are the standard spherical polar coordinates. In terms of $(\zeta,
\bar\zeta)$, the Cartesian coordinates of the point $p^i\in U_n$ are 
\begin{equation}
p^i=P\Bigl(\frac{\bar\zeta+\zeta}{1+\zeta\bar\zeta},{\rm i}\frac{\bar\zeta-
\zeta}{1+\zeta\bar\zeta},\frac{\zeta\bar\zeta-1}{1+\zeta\bar\zeta}\Bigr). 
\label{eq:A.1.2}
\end{equation}
The outward pointing unit normal to ${\cal S}$ at the point $p^i$ is $n^i:=
p^i/P$. This normal is completed to be a basis by the complex vector field 
\begin{equation}
m^i:=\frac{1}{\sqrt{2}}\Bigl(\frac{1-\zeta^2}{1+\zeta\bar\zeta},
  {\rm i}\frac{1+\zeta^2}{1+\zeta\bar\zeta},\frac{2\zeta}{1+\zeta\bar\zeta}
\Bigr) \label{eq:A.1.3}
\end{equation}
and its complex conjugate $\bar m^i$. These are orthogonal to $n^i$, null 
(i.e. $m^im_i=0$), normalized with respect to each other (i.e. $m^i\bar m_i
=1$), and $p^im^j\bar m^k\varepsilon_{ijk}={\rm i}P$ holds. (Recall that, in 
the present paper, the metric on $\mathbb{R}^3$ is chosen to be the 
\emph{positive definite} $\delta_{ij}$, rather than the negative definite 
spatial part of the Minkowski metric $\eta_{ab}:={\rm diag}(1,-1,-1,-1)$, 
where $a,b=0,i$.) In fact, apart from a phase factor, the complex null 
vectors $m^i$ and $\bar m^i$ are uniquely determined by the intrinsic 
complex structure of ${\cal S}$ (see e.g. \cite{HT}). The vector field $m^i$, 
as a differential operator, is given by 
\begin{equation}
m^i\bigl(\frac{\partial}{\partial p^i}\bigr)=\frac{1}{\sqrt{2}P}\bigl(
1+\zeta\bar\zeta\bigr)\Bigl(\frac{\partial}{\partial\bar\zeta}\Bigr). 
\label{eq:A.1.3m}
\end{equation}
Hence, $\zeta$ is a local \emph{anti-holomorphic} coordinate on $U_n$. Also 
in these coordinates, the line element of the metric and the corresponding 
area element on ${\cal S}$ of radius $P$, respectively, are 
\begin{equation}
dh^2=\frac{4P^2}{(1+\zeta\bar\zeta)^2}d\zeta d\bar\zeta, \hskip 20pt 
{\rm d}{\cal S}=\frac{-2{\rm i}P^2}{(1+\zeta\bar\zeta)^2}d\zeta\wedge d
\bar\zeta. \label{eq:A.1.4}
\end{equation}
These are just the metric and area element inherited from the metric and 
volume element of $\mathbb{R}^3$, respectively. There are analogous 
constructions on $U_s:={\cal S}-\{(0,0,-P)\}$, on the 2-sphere minus the 
\emph{south} pole, too; and the structures defined on $U_n$ are related to 
those introduced on $U_s$ smoothly on the overlap $U_n\cap U_s$. 

Considering $\mathbb{R}^3$ to be the $p^0=P$ hyperplane of the Minkowski 
space $\mathbb{R}^{1,3}$ with the Cartesian coordinates $p^a=(p^0,p^i)$ 
and the flat metric $\eta_{ab}$, the 2-sphere ${\cal S}$ is just the 
intersection of the $p^0=P$ hyperplane with the null cone of the origin 
$p^a=0$ in $\mathbb{R}^{1,3}$. Hence, for any $p^i\in{\cal S}$, there is 
a spinor $\pi^A$, the `spinor constituent' of $p^i$, such that $p^i=
\sigma^i_{AA'}\pi^A\bar\pi^{A'}$, and $\pi^A$ is unique only up to the phase 
ambiguity $\pi^A\mapsto\exp({\rm i}\gamma)\pi^A$, $\gamma\in[0,2\pi)$. 
(Note that we lower and raise the small Latin indices by the 
\emph{positive definite} $\delta_{ij}$ and its inverse.) Since $P^2=
\delta_{ij}p^ip^j=-\eta_{ij}\sigma^i_{AA'}\sigma^j_{BB'}\pi^A\bar\pi^{A'}\pi^B
\bar\pi^{B'}=-(\eta_{ab}\sigma^a_{AA'}\sigma^b_{BB'}-\sigma^0_{AA'}\sigma^0
_{BB'})\pi^A\bar\pi^{A'}\pi^B\bar\pi^{B'}=(\sigma^0_{AA'}\pi^A\bar\pi^{A'})^2$, 
the norm of $\pi^A$ with respect to $\sqrt{2}\sigma^0_{AA'}$, i.e. to 
$\sqrt{2}T_{AA'}$ of the previous subsection, is $\sqrt{\sqrt{2}P}$. 
However, $\pi^A$ as a spinor \emph{field} is well defined only on 
${\cal S}$ \emph{minus one point} (see e.g. \cite{HT,EaTod}). In particular, 
on $U_n$, this spinor field, up to a phase, is $\pi^A=(\sqrt{2}P)^{1/2}
o^A$, where $o^A$ is given by (\ref{eq:A.1.1}). Thus $\pi^A$ on $U_n$ and 
the analogous one on $U_s$ are only \emph{locally defined} `spinorial 
coordinates' on ${\cal S}$. 

Using the Newman--Penrose spinor basis $\{o^A,\iota^A\}$, given 
explicitly on $U_n$ by (\ref{eq:A.1.1}) with the choice $O^A=\delta^A_0$ 
and $I^A=\delta^A_1$, it is easy to verify that $m^i\sigma^{AA'}_i=-o^A\bar
\iota^{A'}$ and $p^i\sigma^{AA'}_i=P(\iota^A\bar\iota^{A'}-o^A\bar o^{A'})/
\sqrt{2}$ also hold.

\subsubsection{The line bundles ${\cal O}(-2s)$ over ${\cal S}$}
\label{sub-A.1.3}

One way of defining the complex line bundles ${\cal O}(-2s)$ over ${\cal S}$ 
can be based on the concept of the bundle of totally symmetric N-type 
spinor fields of rank $2\vert s\vert$ on ${\cal S}$: if $s=-\vert s\vert
\leq0$ then these spinor fields are \emph{unprimed} and their principal 
spinor at the point $p^i=\sigma^i_{AA'}\pi^A\bar\pi^{A'}$ is $\pi^A$; and if 
$s=\vert s\vert>0$ then the spinor fields are \emph{primed} and their 
principal spinor at $p^i$ is $\bar\pi^{A'}$. (Recall that e.g. $\lambda^A$ is 
a $2\vert s\vert$-fold principal spinor of the totally symmetric spinor $\phi
^{A_1...A_{2\vert s\vert}}$ if $\phi^{A_1...A_{2\vert s\vert}}\lambda_{A_1}=0$ holds, in 
which case $\phi^{A_1...A_{2\vert s\vert}}$ necessarily has the form $\phi\lambda
^{A_1}\cdots\lambda^{A_{2\vert s\vert}}$ for some $\phi$; and the algebraic type 
of the spinor is called null or of type N, see e.g. \cite{PR,HT}.) Hence, 
e.g. on the domain $U_n$, these spinor fields have the form $\phi
^{A_1...A_{2\vert s\vert}}=\phi\pi^{A_1}...\pi^{A_{2\vert s\vert}}$ and $\chi
^{A'_1...A'_{2\vert s\vert}}=\chi\bar\pi^{A'_1}...\bar\pi^{A'_{2\vert s\vert}}$, where 
$\phi$ and $\chi$ are complex functions on $U_n$. Thus, the fibers of these 
bundles are one complex dimensional, and the line bundle ${\cal O}(-2s)$ is 
just the abstract bundle of these fibers over ${\cal S}$. $U_n$ and $U_s$ 
are local trivialization domains of ${\cal O}(-2s)$, and the functions 
$\phi$ for $s=-\vert s\vert$ (and $\chi$ for $s=\vert s\vert$) are 
\emph{local} cross sections of ${\cal O}(-2s)$ on $U_n$. ${\cal O}(-2s)$ is 
globally trivializable precisely when $s=0$. 

The phase ambiguity $\pi^A\mapsto\exp({\rm i}\gamma)\pi^A$ in the 
principal spinor yields the ambiguity $\phi\mapsto\exp(-2{\rm i}\vert s\vert
\gamma)\phi$, where $\gamma$ is an arbitrary $[0,2\pi)$-valued locally 
defined function on ${\cal S}$. The analogous ambiguity in the function
$\chi$ is $\chi\mapsto\exp(2{\rm i}s\gamma)\chi$. Therefore, despite 
this ambiguity, the Hermitian scalar product of any two cross sections, 
representing e.g. the spinor fields $\phi^{A_1...A_{2\vert s\vert}}$ and $\psi
^{A_1...A_{2\vert s\vert}}$ on ${\cal S}$ and given by 
\begin{equation}
\langle\phi^{A_1...A_{2\vert s\vert}},\psi^{A_1...A_{2\vert s\vert}}\rangle_s
:=\int_{\cal S}\bar\phi\psi{\rm d}{\cal S}, \label{eq:A.1.5}
\end{equation}
is well defined. The space of the square-integrable cross sections of 
${\cal O}(-2s)$ is a Hilbert space, denoted by ${\cal H}_s$. An 
alternative, and perhaps more familiar form of this scalar product can be 
given in terms of the spinor fields themselves. To rewrite (\ref{eq:A.1.5}) 
in this form, let us recall from subsection \ref{sub-A.1.2} that $\sigma^0
_{AA'}=(O_A\bar O_{A'}+I_A\bar I_{A'})/\sqrt{2}=(o_A\bar o_{A'}+\iota_A\bar
\iota_{A'})/\sqrt{2}$. Using this, the integrand of (\ref{eq:A.1.5}) can be 
rewritten as 
\begin{eqnarray}
\bar\phi\psi\!\!\!\!&=\!\!\!\!&(\sqrt{2}P)^{-2\vert s\vert}\bar\iota_{A'_1}
  \cdots\bar\iota_{A'_{2\vert s\vert}}\bar\phi^{A'_1...A'_{2\vert s\vert}}\iota_{A_1}
  \cdots\iota_{A_{2\vert s\vert}}\psi^{A_1...A_{2\vert s\vert}} \nonumber \\
\!\!\!\!&=\!\!\!\!&P^{-2\vert s\vert}\sigma^0_{A_1A'_1}\cdots\sigma^0
  _{A_{2\vert s\vert}A'_{2\vert s\vert}}\bar\phi^{A'_1...A'_{2\vert s\vert}}\psi
  ^{A_1...A_{2\vert s\vert}}. \label{eq:A.1.6}
\end{eqnarray}
If we think of $\sigma^{AA'}_0$ as the components of the spinor form 
of the timelike vector of the orthonormal vector basis in a Lorentzian 
vector space whose spatial vectors span $\mathbb{R}^3$, then by 
(\ref{eq:A.1.6}) the scalar product (\ref{eq:A.1.5}) is just $(\sqrt{2}P)
^{-2\vert s\vert}$ times the scalar product that is analogous to the familiar, 
standard $L_2$-scalar product of two spinor fields on the mass shell in 
Poincar\'e-invariant quantum theory (see \cite{StWi}). 

Recalling that a scalar $\phi$ is said to have the spin weight $\frac{1}{2}
(p-q)$ if under the rescaling $\{o^A,\iota^A\}\mapsto\{\lambda o^A,
\lambda^{-1}\iota^A\}$, where $\lambda$ is any nowhere vanishing complex 
function on the domain of the spin frame, the scalar $\phi$ transforms as 
$\phi\mapsto\lambda^p\bar\lambda^q\phi$ (see e.g. \cite{PR,HT}), we can see 
that ${\cal O}(-2s)$ is just the bundle of spin weighted scalars of weight 
$s$ on ${\cal S}$. In particular, the components of the vectors $m^i$ and 
$\bar m^i$ are of types $(1,-1)$ and $(-1,1)$, respectively, while those of 
$p^i$ are sums of a $(1,1)$ and a $(-1,-1)$ type scalar. Thus, the spin 
weight of $m^i$, $\bar m^i$ and $p^i$ is $1$, $-1$ and $0$, respectively; 
and the spin weight of the spherical harmonics ${}_sY_{j,m}$, defined by 
(\ref{eq:A.1.1Y}), is $s$. 

If $\delta_i$ denotes the (Cartesian components of the) covariant derivative 
operator of the induced Levi-Civita connection acting on the spinor fields 
on ${\cal S}$, then the edth and edth-prime operators of Newman and Penrose 
\cite{NP} are defined as the covariant directional derivative operators $m^i
\delta_i$ and $\bar m^i\delta_i$, respectively, acting on the cross sections 
$\phi$ of the line bundles ${\cal O}(-2s)$. Explicitly, if $\phi
_{A_1...A_{2\vert s\vert}}=\phi\pi_{A_1}\cdots\pi_{A_{2\vert s\vert}}$, then ${\edth}_s\phi
:=m^i\delta_i((\sqrt{2}P)^{-\vert s\vert}\phi_{A_1...A_{2\vert s\vert}})\iota^{A_1}...
\iota^{A_{2\vert s\vert}}$ and ${\edth}'_s\phi:=\bar m^i\delta_i((\sqrt{2}P)
^{-\vert s\vert}\phi_{A_1...A_{2\vert s\vert}})\iota^{A_1}...\iota^{A_{2\vert s\vert}}$ (see 
also \cite{PR,HT,EaTod,Goetal}). ${\edth}_s$ and ${\edth}'_s$ acting on 
cross sections of ${\cal O}(-2s)$ with $s=\vert s\vert$ are defined 
analogously. ${\edth}_s$ increases, and ${\edth}'_s$ decreases the spin 
weight by one. In the complex stereographic coordinates on $U_n$, the 
explicit form of these operators, acting on a function $\phi$ of spin weight 
$s$, is 
\begin{equation}
{\edth}_s\phi=\frac{1}{\sqrt{2}P}\Bigl(\bigl(1+\zeta\bar\zeta\bigr)
\frac{\partial\phi}{\partial\bar\zeta}+s\zeta\phi\Bigr), \hskip 20pt
{\edth}'_s\phi=\frac{1}{\sqrt{2}P}\Bigl(\bigl(1+\zeta\bar\zeta\bigr)
\frac{\partial\phi}{\partial\zeta}-s\bar\zeta\phi\Bigr). \label{eq:A.1.7}
\end{equation}
If no confusion arises, simply we write ${\edth}$ and ${\edth}'$ instead of 
${\edth}_s$ and ${\edth}'_s$. These operators link the spinors $o^A$ and 
$\iota^A$: ${\edth}o^A=0$, ${\edth}'o^A=\iota^A/(\sqrt{2}P)$, ${\edth}\iota^A
=-o^A/(\sqrt{2}P)$ and ${\edth}'\iota^A=0$; which imply ${\edth}p^i=m^i$, 
${\edth}m^i=0$ and ${\edth}\bar m^i=-p^i/P^2$. Using these formulae, it is 
easy to see that 
\begin{eqnarray}
&{}&{\edth}z(j,s)_{A_1...A_{2j}}=-\frac{1}{\sqrt{2}P}(j-s)z(j,s+1)_{A_1...A_{2j}}, 
\label{eq:A.1.7a} \\
&{}&{\edth}'z(j,s)_{A_1...A_{2j}}=\frac{1}{\sqrt{2}P}(j+s)z(j,s-1)_{A_1...A_{2j}},
\label{eq:A.1.7b}
\end{eqnarray}
where the spinor $z(j,s)_{A_1...A_{2j}}$ was introduced in Appendix 
\ref{sub-A.1.1}. Thus, while the ladder operators ${\bf J}_\pm:={\bf J}_1\pm
{\rm i}{\bf J}_2$ change the quantum number $m$ of the canonical angular 
momentum basis vectors $\vert j,m\rangle$ by one, ${\edth}$ increases and 
${\edth}'$ decreases the `quantum number' $s$ of $z(j,s)_{A_1...A_{2j}}$ by one. 
Hence, ${\edth}$ and ${\edth}'$ play the role analogous to the ladder 
operators. 

The spin weighted spherical harmonics ${}_sY_{j,m}$ can also be defined (up 
to phase and normalization) by the pair of equations ${\edth}\,{}_sY_{j,m}=
-\frac{1}{\sqrt{2}P}\sqrt{(j+s+1)(j-s)}\,{}_{s+1}Y_{j,m}$ and ${\edth}'\,{}_s
Y_{j,m}=\frac{1}{\sqrt{2}P}\sqrt{(j-s+1)(j+s)}\,{}_{s-1}Y_{j,m}$. These imply 
that the harmonics are eigenfunctions of ${\edth}{\edth}'$ and ${\edth}'
{\edth}$, and also of the Laplacian: $({\edth}{\edth}'+{\edth}'{\edth}){}_s
Y_{j,m}=-P^{-2}(j(j+1)-s^2){}_sY_{j,m}$. (For more details, see 
\cite{NP,Goetal,PR,HT}.) 

For the general, abstract definition of the bundles ${\cal O}(-2s)$ over 
complex projective spaces, see e.g. \cite{PR,HT,EaTod}. For their introduction 
and a discussion of some of the global topological properties of the operators 
${\edth}$ and ${\edth}'$ on closed metric 2-surfaces even with any genus, see 
\cite{FrSz}.

\subsubsection{The angular momentum operators on ${\cal O}(-2s)$}
\label{sub-A.1.4}

In this subsection, we determine the specific form of the angular momentum 
operators acting on spin weighted functions. This geometrical form is more 
natural than the usual one when the algebra of the basic quantum observables, 
$su(2)$, is considered to be a subalgebra of the Euclidean algebra $e(3)$, 
which will be considered in \cite{Sz2}. The form of this representation is 
similar to that of the generators of the Poincar\'e algebra of relativistic 
quantum systems \cite{StWi}, too. 

The action of the group $SU(2)$ on $\mathbb{R}^3$ is defined by $p^i\mapsto
\Lambda^i{}_j(A)p^j$, where $\Lambda^i{}_j(A):=-\sigma^i_{AA'}A^A{}_B
\bar A^{A'}{}_{B'}\sigma^{BB'}_j$, in which $A^A{}_B\in SU(2)$ and over-bar 
denotes complex conjugation. (The $(-)$ sign in the expression of $\Lambda
^i{}_j(A)$ is due to our convention that we lower and raise the small Latin 
indices by the \emph{positive definite} $\delta_{ij}$ and its inverse.) Thus 
$SU(2)$ is the (universal covering group of the) group of those 
\emph{isometries} of the \emph{flat Riemannian 3-manifold} $(\mathbb{R}^3,
\delta_{ij})$ that leave the origin $p^i=0$ fixed. The surfaces of 
transitivity of $SU(2)$ are the 2-spheres ${\cal S}$ with radius $P>0$ and 
the origin $p^i=0$. The latter case is uninteresting for us in the present 
paper, because that yields the trivial representation. 

One way of determining the irreducible representations of $SU(2)$ is by 
means of the method of induced representations. In this way, first we 
should find the representations of the stabilizer subgroup for a point 
$\mathring{p}^i\in{\cal S}$ in $SU(2)$. This is $U(1)\subset SU(2)$, and, 
by Schur's lemma, all of its irreducible representations are one-dimensional, 
and these are labeled by $s=0,\pm\frac{1}{2},\pm1,...$. If $\mathring{\pi}
{}^A$ is the spinor constituent of $\mathring{p}^i$ (see Appendix 
\ref{sub-A.1.2}), then this one-dimensional representation space is chosen 
to be spanned by the spinor of the form $\mathring{\pi}{}^{A_1}\cdots
\mathring{\pi}{}^{A_{2\vert s\vert}}$ if $s=-\vert s\vert\leq0$, and $\bar
{\mathring{\pi}}{}^{A'_1}\cdots\bar{\mathring{\pi}}{}^{A'_{2s}}$ if $s=
\vert s\vert>0$. The next step is the generation of the representation 
space for the whole group $SU(2)$ from this one dimensional space by the 
elements of $SU(2)$ that do not leave $\mathring{p}^i$ fixed. 

Geometrically, the above method (by using the group action from $\mathring{p}
^i$) is the construction of the bundle of totally symmetric unprimed N-type 
spinors $\phi^{A_1...A_{2\vert s\vert}}$ on ${\cal S}$ if $s=-\vert s\vert$, and 
of the totally symmetric primed N-type spinors $\chi^{A'_1...A'_{2s}}$ on 
${\cal S}$ if $s=\vert s\vert$. (Clearly, these are equivalent to the line 
bundles ${\cal O}(-2s)$ with the corresponding $s$.) The $2\vert s\vert$-fold 
principal spinor of them at the point $p^i=\sigma^i_{AA'}\pi^A\bar\pi^{A'}$ is 
$\pi^A$ and $\bar\pi^{A'}$, respectively. 

To determine the explicit form of the representation of $SU(2)$ by operators 
${\bf U}(A)$ acting on the spinor fields, let us recall that the rotation 
Killing vectors $k^i_{mn}$ are tangent to ${\cal S}$ and also generate its 
isometries. Then the action of $SU(2)$ e.g. on any $\phi^{A_1...A_{2\vert s\vert}}$ 
is defined by $({\bf U}(A)\phi)^{A_1...A_{2\vert s\vert}}(p^i):=A^{A_1}{}_{B_1}
\cdots A^{A_{2\vert s\vert}}{}_{B_{2\vert s\vert}}\phi^{B_1...B_{2\vert s\vert}}(\Lambda(
A^{-1})^i{}_jp^j)$. In particular, apart from a phase factor, $({\bf U}(A)\pi)
^A$ is just $\pi^A$, i.e. the spinor constituent of the position vector field 
is $SU(2)$-invariant up to a phase. Hence, any $2\vert s\vert$-rank spinor 
field that belongs to the carrier space of an irreducible representation 
of $SU(2)$ is necessarily $N$-type with $\pi^A$ and $\bar\pi^{A'}$ as its 
$2\vert s\vert$-fold principal spinor for $s=-\vert s\vert$ and $s=\vert s
\vert$, respectively. Note that the function $\phi$ appearing in $\phi
^{A_1...A_{2\vert s\vert}}=\phi\pi^{A_1}\cdots\pi^{A_{2\vert s\vert}}$ has spin weight 
$s$. Using (\ref{eq:A.1.6}), it is straightforward to check that the operator 
${\bf U}(A)$ is unitary with respect to the scalar product (\ref{eq:A.1.5}). 
As we will see in \cite{Sz2}, the $L_2$ space ${\cal H}_s$ of the $N$-type 
spinor fields with given $s$ provides the carrier space of an 
\emph{irreducible} representation for the $E(3)$ group, but it is \emph{not} 
irreducible for the $SU(2)$. 

In this representation, still for $s=-\vert s\vert\leq0$, the operators 
${\bf J}_{ij}:=\varepsilon_{ijk}{\bf J}^k$ are defined to be the densely 
defined self-adjoint generators of these transformations: let $A^A{}_B(u)$ 
be a 1-parameter subgroup in $SU(2)$ generated by $\lambda^A{}_B\in su(2)$. 
Then its trajectories on ${\cal S}$ are necessarily integral curves of 
some rotation Killing vector field, say of $M^{mn}k^i_{mn}$ for some 
constants $M^{mn}=-M^{nm}$. Then ${\bf J}_{ij}$ is defined by $({\rm i}/\hbar)
M^{ij}{\bf J}_{ij}\phi^{A_1...}:=\frac{\rm d}{{\rm d}u}(({\bf U}(A(u))\phi)
^{A_1...})\vert_{u=0}$. This is just minus the Lie derivative of the spinor 
field along the Killing vector $M^{mn}k^i_{mn}$ (see e.g. \cite{PR,HT}). Here 
the limit in the definition of the derivative is meant in the strong topology 
of ${\cal H}_s$. To evaluate this, let us calculate the tangent of the 
trajectories $\Lambda^i{}_j(A(u))p^j$ at $p^i$. This is $k^i=-\sigma^i
_{AA'}(\lambda^A{}_B\delta^{A'}_{B'}+\delta^A_B\bar\lambda^{A'}{}_{B'})\sigma
^{BB'}_jp^j$, which must coincide with $M^{mn}k^i_{mn}=2M^i{}_jp^j$. Since 
$\lambda^A{}_B\in su(2)$, its complex conjugate is not independent of 
$\lambda^A{}_B\in su(2)$, because $\lambda_{AB}=2\bar\lambda^{A'B'}\sigma^0
_{A'A}\sigma^0_{B'B}$ holds. Thus, introducing the standard $SU(2)$ Pauli 
matrices $\sigma^A_i{}_B$ (including the factor $1/\sqrt{2}$) as the unitary 
spinor form of the three non-trivial $SL(2,\mathbb{C})$ Pauli matrices (i.e. 
which are given explicitly by $\sigma^A_i{}_B=\delta_{ij}\varepsilon^{AC}
\sigma^j{}_{CB'}\sqrt{2}\sigma^{0B'}{}_B$), we can express $\lambda^A{}_B$ in 
terms of $M^{ij}$: 
\begin{equation}
\lambda^A{}_B=-\frac{\rm i}{\sqrt{2}}M^{ij}\varepsilon_{ij}{}^k
\sigma^A_k{}_B. \label{eq:A.1.8*}
\end{equation}
In deriving this formula we also used the identity 
\begin{equation*}
\sigma^A_i{}_B\sigma_j^B{}_C=-\frac{\rm i}{\sqrt{2}}\varepsilon_{ij}{}^k
\sigma^A_k{}_C+\frac{1}{2}\delta_{ij}\delta^A_C.
\end{equation*}
Now, using (\ref{eq:A.1.8*}), we are able to evaluate the equation defining 
the operator ${\bf J}_{ij}$. We find that, still for $s=-\vert s\vert$, 
\begin{equation}
{\bf J}_{ij}\phi^{A_1...A_{\vert s\vert}}={\rm i}\hbar\Bigl(p_j\frac{\partial}
{\partial p^i}-p_i\frac{\partial}{\partial p^j}\Bigr)\phi^{A_1...A_{\vert s\vert}}
+\sqrt{2}\hbar\, s\,\varepsilon_{ij}{}^k\sigma_k^{(A_1}{}_{(B_1}\delta^{A_2}
_{B_2}\cdots\delta^{A_{2\vert s\vert})}_{B_{2\vert s\vert})}\phi^{B_1...B_{\vert s\vert}}.
\label{eq:A.1.8}
\end{equation}
Thus, ${\bf J}_{ij}$ is ${\rm i}\hbar$-times the Lie derivative operator 
along the Killing vector $k^l_{ij}$, and hence is well defined on the 
\emph{dense subspace} of the smooth spinor fields in ${\cal H}_s$. Then it is 
a straightforward calculation to check that these operators do, indeed, 
satisfy the defining commutation relations of $su(2)$ on the appropriate 
dense subspaces, i.e. provide a representation of the Lie algebra of the 
rotation Killing fields. Repeating the analogous analysis for $s=\vert s
\vert$, we obtain the same expression (\ref{eq:A.1.8}) for ${\bf J}_{ij}$. 

Next, using the form ${\bf J}_i=\varepsilon_i{}^{jk}{\bf J}_{jk}/2$ of the 
angular momentum operator, we determine its contraction with the basis 
vectors $p^i$, $m^i$ and $\bar m^i$. First, recalling from Appendix 
\ref{sub-A.1.2} that $p^i\sigma_i^{AA'}=P(\iota^A\bar\iota^{A'}-o^A\bar o
^{A'})/\sqrt{2}$, for its unitary spinor form we obtain 
\begin{equation*}
p^i\sigma^A_i{}_B:=p^i\sigma^A_i{}_{A'}\sqrt{2}\sigma^{A'}_0{}_B=-\frac{P}
{\sqrt{2}}\bigl(\iota^A\bar\iota^{A'}-o^A\bar o^{A'}\bigr)\bigl(\bar o_{A'}
o_A+\bar\iota_{A'}\iota_A\bigr)=-\frac{P}{\sqrt{2}}\bigl(\iota^Ao_B+o^A
\iota_B\bigr).
\end{equation*}
Since on the domain $U_n$ the spinor field has the form $\phi
^{A_1...A_{2\vert s\vert}}=\phi\pi^{A_1}\cdots\pi^{A_{2\vert s\vert}}$, we find that  
\begin{equation}
p^i{\bf J}_i\phi^{A_1...A_{2\vert s\vert}}=sP\hbar\phi^{A_1...A_{2\vert s\vert}}.
\label{eq:A.1.9a}
\end{equation}
It might be worth noting that in the \emph{unitary, irreducible} 
representation of the Euclidean group $E(3)$ exactly the same expression 
emerges as the analog of the Pauli--Lubanski spin operator \cite{Sz2}, which 
is one of the two Casimir operators of $e(3)$. 

Similarly, one can show that $m^i\sigma^A_i{}_B=o^Ao_B=\pi^A\pi_B/(\sqrt{2}
P)$ and $\bar m^i\sigma^A_i{}_B=-\iota^A\iota_B$. Then, using $m^ip^j
\varepsilon_{ijk}=-p^im^j\bar m^l\varepsilon_{ijl}m_k=-{\rm i}Pm_k$ and $\bar m
^ip^j\varepsilon_{ijk}={\rm i}P\bar m_k$ (see Appendix \ref{sub-A.1.2}), as 
well as the definition of the ${\edth}$ and ${\edth}'$ operators (Appendix 
\ref{sub-A.1.3}), we find 
\begin{eqnarray}
&{}&m^i{\bf J}_i\phi^{A_1...A_{2\vert s\vert}}=-P\hbar\bigl({\edth}\phi\bigr)\pi
  ^{A_1}\cdots\pi^{A_{2\vert s\vert}}. \label{eq:A.1.9b} \\
&{}&\bar m^i{\bf J}_i\phi^{A_1...A_{2\vert s\vert}}=P\hbar\bigl({\edth}'\phi\bigr)
  \pi^{A_1}\cdots\pi^{A_{2\vert s\vert}}. \label{eq:A.1.9c}
\end{eqnarray}
Hence, by (\ref{eq:A.1.9a})-(\ref{eq:A.1.9c}), 
\begin{eqnarray*}
{\bf J}_i\phi^{A_1...A_{2\vert s\vert}}\!\!\!\!&=\!\!\!\!&\bigl(m_i\bar m^j+\bar m
  _im^j+\frac{1}{P^2}p_ip^j\bigr){\bf J}_j\phi^{A_1...A_{2\vert s\vert}} \\
\!\!\!\!&=\!\!\!\!&P\hbar\Bigl(m_i{\edth}'\phi-\bar m_i{\edth}\phi+s\frac{p_i}
  {P^2}\phi\Bigr)\pi^{A_1}\cdots\pi^{A_{2\vert s\vert}}.
\end{eqnarray*}
Thus, ${\bf J}_i$ preserves the algebraic, viz. the N-type of the spinor 
spinor fields. Therefore, defining the action of ${\bf J}_i$ on the spin 
weighted function $\phi$ with spin weight $s$ simply by ${\bf J}_i\phi:=
(\sqrt{2}P)^{-\vert s\vert}({\bf J}_i\phi_{A_1...A_{2\vert s\vert}})\iota^{A_1}\cdots
\iota^{A_{2\vert s\vert}}$, we obtain 
\begin{equation}
{\bf J}_i\phi=P\hbar\Bigl(m_i{\edth}'\phi-\bar m_i{\edth}\phi+s\frac{n_i}{P}
\phi\Bigr). \label{eq:A.1.10}
\end{equation}
Hence, the operators ${\bf J}_i$ map cross sections of ${\cal O}(-2s)$ into 
cross sections of ${\cal O}(-2s)$. In $E(3)$ invariant elementary quantum 
mechanical systems the same expression will be interpreted in \cite{Sz2} 
as the decomposition of the total angular momentum into its orbital and 
spin parts, where $P$ and $s$ are fixed by the two Casimir operators of 
$e(3)$. 

By (\ref{eq:A.1.10}) it is easy to compute the only Casimir operator 
${\bf J}_i{\bf J}^i$ of the $su(2)$ algebra. It is 
\begin{equation}
{\bf J}_i{\bf J}^i\phi=\hbar^2\Bigl(-P^2\bigl({\edth}{\edth}'+{\edth}'{\edth}
\bigr)\phi+s^2\phi\Bigr); \label{eq:A.1.11}
\end{equation}
where, as we noted at the end of Appendix \ref{sub-A.1.3}, ${\edth}{\edth}'
+{\edth}'{\edth}$ is just the Laplace operator on ${\cal S}$, and there we 
determined its spectrum and eigenfunctions. Using these, we find that the 
eigenvalues of ${\bf J}_i{\bf J}^i$ are $j(j+1)\hbar^2$, $j=\vert s\vert,
\vert s\vert+1,...$, and the corresponding eigenfunctions are the spin 
weighted spherical harmonics ${}_sY_{j,m}$, $m=-j,-j+1,...,j$. To find the 
${\bf J}_i$-invariant subspaces of ${\cal H}_s$, it is enough to recall that 
${\bf J}_i$ and ${\bf J}_k{\bf J}^k$ commute, and hence we find that, for 
fixed $s$ and $j$, the $SU(2)$-irreducible subspace ${\cal H}_{s,j}\subset
{\cal H}_s$ is spanned just by the spin weighted spherical harmonics ${}_s
Y_{j,m}$, $m=-j,-j+1,...,j$. Thus all these irreducible subspaces are $2j+1$ 
dimensional, and the edth and edth-prime operators, which are adjoint of 
each other, are maps between them: ${\edth}:{\cal H}_{s,j}\to{\cal H}_{s+1,j}$ 
and ${\edth}':{\cal H}_{s,j}\to{\cal H}_{s-1,j}$. This structure of the 
$SU(2)$-irreducible subspaces together with the structure (\ref{eq:A.1.10}) 
of the operators ${\bf J}_i$ and its interpretation in $E(3)$-invariant 
systems justify the interpretation of $s$ as the \emph{intrinsic spin} of 
the system, even though it is not the value of any Casimir operator of 
$su(2)$.

\subsubsection{A quantum information theoretical interpretation of the 
spinorial notions} 
\label{sub-A.1.5}

The spin-space $\mathbb{S}_A$ could be interpreted as the Hilbert space of 
the states of a two-states quantum system (`qubit'); while $\mathbb{S}
_{(A_1...A_{2j})}$, the symmetrized $2j$-fold tensor product of the spin-space 
with itself, as the Hilbert space of the states of $2j$ identical and 
indistinguishable (`bosonic') qubits. The basis vectors $O_A$ and $I_A$ in 
$\mathbb{S}_A$ are usually denoted in quantum information theory by $\vert
0\rangle$ and $\vert1\rangle$, respectively. The basis vector $Z(j,m)
_{A_1...A_{2j}}$, $m=-j,-j+1,...,j$, in $\mathbb{S}_{(A_1...A_{2j})}$ corresponds 
to the one obtained by complete symmetrization of the tensor product basis 
vectors in $\mathbb{S}_{A_1}\otimes\cdots\otimes\mathbb{S}_{A_{2j}}$ according 
to 
\begin{equation*}
\vert(0,\cdots,0,1,\cdots,1)\rangle:=\frac{1}{(2j)!}\sum_\pi\vert\pi\bigl(
0,\cdots,0,1,\cdots,1\bigr)\rangle.
\end{equation*}
Here $\pi(0,\cdots,0,1,\cdots,1)$ denotes a permutation of the array $(0,
\cdots,0,1,\cdots,1)$, the summation is on all the permutations, and the 
number of $0$'s in each term is $(j-m)$ and that of $1$'s is $(j+m)$. This 
vector is usually denoted by $\vert j,m\rangle$. In particular, $\vert j,
-j\rangle=\vert0,\cdots,0\rangle$ and $\vert j,j\rangle=\vert1,\cdots,1
\rangle$. 

As is well known (see e.g. \cite{NiChu}), the set of the pure quantum 
states of a qubit is homeomorphic to the unit sphere $S^2$, the so-called 
Bloch sphere. Parameterizing this sphere by the familiar angle coordinates 
$(\theta,\varphi)$, the coordinates $(\theta',\varphi')$ of its anti-podal 
point on $S^2$ are $\theta'=\pi-\theta$, $\varphi'=\varphi+\pi$. Then, in 
the basis $\{\vert0\rangle,\vert1\rangle\}=\{O^A,I^A\}$, the quantum states 
corresponding to these points of the Bloch sphere, $\vert\psi\rangle=\psi^A$ 
and $\vert\psi'\rangle=\psi'{}^A$, are 
\begin{eqnarray}
&{}&\psi^A:=\cos\frac{\theta}{2}O^A+\exp({\rm i}\varphi)\sin\frac{\theta}{2}
  I^A, \label{eq:A.1.5.1a}\\
&{}&\psi'{}^A:=\cos\frac{\theta'}{2}O^A+\exp({\rm i}\varphi')\sin
  \frac{\theta'}{2}I^A=-\sin\frac{\theta}{2}O^A+\exp({\rm i}\varphi)\cos
  \frac{\theta}{2}I^A. \label{eq:A.1.5.1b}
\end{eqnarray}
Clearly, $\{\psi^A,\psi'{}^A\}$ is a basis in $\mathbb{S}^A$, this is 
orthonormal with respect to the Hermitian scalar product $\sqrt{2}T_{AA'}$, 
but it is \emph{not} a normalized spin frame: $\varepsilon_{AB}\psi^A
\psi'{}^B=\psi_A\psi'{}^A=\exp({\rm i}\varphi)\not=1$. In fact, the 
determinant of the matrix of the basis transformation $\{O^A,I^A\}\mapsto
\{\psi^A,\psi'{}^A\}$ given by (\ref{eq:A.1.5.1a})--(\ref{eq:A.1.5.1b}) is 
$\exp({\rm i}\varphi)$, rather than $1$. Hence this matrix is \emph{not} 
an $SU(2)$ spin transformation. 

The vectors $o^A$ and $\iota^A$ of the Newman--Penrose spin frame on 
${\cal S}$, given by (\ref{eq:A.1.1}), can also be interpreted as quantum 
states of the qubit. In fact, identifying the point $(\theta,\varphi)$ of 
the Bloch sphere with the point with the \emph{complex conjugate} 
stereographic coordinate $\eta:=\bar\zeta$ on ${\cal S}$ according to $\eta
=\exp({\rm i}\varphi)\cot(\theta/2)$, the vectors of the Newman--Penrose 
spin frame at the point $(\zeta,\bar\zeta)\in{\cal S}$ take the form 
\begin{eqnarray}
&{}&o^A=\frac{-{\rm i}}{\sqrt{1+\zeta\bar\zeta}}\Bigl(\zeta O^A+I^A\Bigr)=
  -{\rm i}\exp(-{\rm i}\varphi)\Bigl(\cos\frac{\theta}{2}O^A+\exp({\rm i}
  \varphi)\sin\frac{\theta}{2}I^A\Bigr), \label{eq:A.1.5.2a}\\
&{}&\iota^A=\frac{-{\rm i}}{\sqrt{1+\zeta\bar\zeta}}\Bigl(O^A-\bar\zeta I^A
  \Bigr)= {\rm i}\Bigl(-\sin\frac{\theta}{2}O^A+\exp({\rm i}\varphi)\cos
  \frac{\theta}{2}I^A\Bigr). \label{eq:A.1.5.2b}
\end{eqnarray}
Thus, $o^A$ is proportional to $\psi^A$ and $\iota^A$ to $\psi'{}^A$ above, 
the factors of proportionality are only pure phases, but these factors are 
\emph{different}. Since $\{o^A,\iota^A\}$ is connected to the basis $\{O^A,
I^A\}$ by an $SU(2)$ basis transformation, it is slightly more natural to 
use the Newman--Penrose basis than $\{\psi^A,\psi'{}^A\}$. 

It is well known \cite{Rad,Kl} that, using the angular momentum ladder 
operators ${\bf J}_\pm:=({\bf J}_1\pm{\rm i}{\bf J}_2)$, for any fixed $j$ 
and $w\in\mathbb{C}$ the states 
\begin{equation*}
\vert w\rangle_\pm:=\frac{1}{(1+\vert w\vert^2)^j}\exp\bigl(\frac{w}{\hbar}
{\bf J}_\pm\bigr)\vert j,\mp j\rangle=\frac{1}{(1+\vert w\vert^2)^j}\sum
_{k=0}^{2j}w^k\sqrt{{2j}\choose{k}}\vert j,\mp(j-k)\rangle,
\end{equation*}
the so-called $SU(2)$ coherent states, are analogous to the canonical 
coherent states of Heisenberg systems; and the latter are usually 
considered to be the `most classical states' of these systems. 

States analogous to $\vert w\rangle_\pm$ were introduced in \cite{Rad}, 
which are based on states of the form $\psi_{A_1}\cdots\psi_{A_{2j}}$ or 
$\psi'_{A_1}\cdots\psi'_{A_{2j}}$ above, instead of $\vert j,\pm j\rangle$. 
(In quantum information theory, these states are usually denoted by $\vert
\psi\rangle^{\otimes2j}$ and $\vert\psi'\rangle^{\otimes2j}$, respectively.) 
Now we show how these analogous states can be introduced in a quite simple 
way, using the ${\edth}$ and ${\edth}'$ operators of Appendix 
\ref{sub-A.1.3}. Recalling from Appendix \ref{sub-A.1.3} how ${\edth}$ 
and ${\edth}'$ act on the spinor field $z(j,s)_{A_1...A_{2j}}$ (see equations 
(\ref{eq:A.1.7a})-(\ref{eq:A.1.7b})), it is easy to see that, for any $k=
0,1,...,2j$, 
\begin{eqnarray*}
&{}&{\edth}^kz(j,-j)_{A_1...A_{2j}}=\bigl(\frac{-1}{\sqrt{2}P}\bigr)^kk!
  {{2j}\choose{k}}z(j,-j+k)_{A_1...A_{2j}}, \label{eq:A.1.5.5a} \\
&{}&{\edth}'{}^kz(j,j)_{A_1...A_{2j}}=\bigl(\frac{1}{\sqrt{2}P}\bigr)^kk!
  {{2j}\choose{k}}z(j,j-k)_{A_1...A_{2j}} \label{eq:A.1.5.5b}
\end{eqnarray*}
hold. Hence, for any $w\in\mathbb{C}$, it is straightforward to show that 
\begin{eqnarray}
\vert w\rangle\rangle_+:=\!\!\!\!&{}\!\!\!\!&\frac{1}{(1+\vert w\vert^2)^j}
  \exp\Bigl(-w\sqrt{2}P\,{\edth}\Bigr)z(j,-j)_{A_1...A_{2j}}= \nonumber \\
=\!\!\!\!&{}\!\!\!\!&\frac{1}{(1+\vert w\vert^2)^j}\sum_{k=0}^{2j}w^k
  {{2j}\choose{k}}z(j,-j+k)_{A_1...A_{2j}},\label{eq:A.1.5.4a} \\
\vert w\rangle\rangle_-:=\!\!\!\!&{}\!\!\!\!&\frac{1}{(1+\vert w\vert^2)^j}
  \exp\Bigl(w\sqrt{2}P\,{\edth}'\Bigr)z(j,j)_{A_1...A_{2j}}= \nonumber \\
=\!\!\!\!&{}\!\!\!\!&\frac{1}{(1+\vert w\vert^2)^j}\sum_{k=0}^{2j}w^k
  {{2j}\choose{k}}z(j,j-k)_{A_1...A_{2j}}. \label{eq:A.1.5.4b}
\end{eqnarray}
It is a simple calculation to demonstrate that the states $\vert w\rangle
\rangle_+$ and $\vert w\rangle\rangle_-$ have unit norm. $\vert w\rangle
\rangle_+$ is orthogonal to $\vert w'\rangle\rangle_+$, and $\vert w\rangle
\rangle_-$ to $\vert w'\rangle\rangle_-$, precisely when $w'=-1/\bar w$ 
(i.e. when $w'$ and $w$ are the complex stereographic coordinates of points 
of $S^2$ that are anti-podal of each other); and $\vert w\rangle\rangle_+$ 
is orthogonal to $\vert w'\rangle\rangle_-$ precisely when $w'=-\bar w$. A 
key property of the coherent states is well known to be that they provide 
the partition of unity. Since the spinors $\vert z(j,s)\rangle=z(j,s)
_{A_1...A_{2j}}$, $s=-j,-j+1,...,j$, form a basis in $\mathbb{S}_{(A_1...A_{2j})}$, 
this key property is equivalent to 
\begin{equation}
z(j,s)_{A_1...A_{2j}}=\frac{1}{\pi}\int_{\mathbb{C}}\vert w\rangle\rangle\langle
\langle w\vert z(j,s)\rangle\,{\tt m}\,{\rm d}^2w \label{eq:A.1.5.5}
\end{equation}
for some weight function ${\tt m}={\tt m}(w)$, which may depend on $j$. 
Here $\vert w\rangle\rangle$ is any of $\vert w\rangle\rangle_\pm$, and the 
integral measure on $\mathbb{C}$ is the natural one: ${\rm d}^2w=\rho\,
{\rm d}\rho\,{\rm d}\chi$, where $w$ is represented by its Euler form $w=
\rho\exp({\rm i}\chi)$. A direct calculation does, in fact, show that, with 
the weight function ${\tt m}=(2j+1)(1+\vert w\vert^2)^{-2}$ given in 
\cite{Rad}, (\ref{eq:A.1.5.5}) holds true.


\subsection{The explicit form of the eigenfunctions $W_{s,j,m}$ and 
the asymptotics of the standard deviations}
\label{sub-A.2}

\subsubsection{The explicit form of the eigenfunctions $\phi_{s,j,m}$}
\label{sub-A.2.1}

The strategy of the determination of the explicit coordinate form of 
$W_{s,j,m}$ and $W_{s,j}$ on ${\cal S}$ (and in particular the factors of 
normalization, $N_{s,j,m}$ and $N_{s,j}$) is just that for the spin weighted 
spherical harmonics ${}_sY_{j,m}$ given in \cite{PR}: the eigenfunction 
\begin{equation}
\phi_{s,j,m}=\mu_{(A_1}\cdots\mu_{A_{j-m}}\nu_{A_{j-m+1}}\cdots\nu_{A_{2j})}o^{A_1}\cdots o
^{A_{j+s}}\iota^{A_{j+s+1}}\cdots\iota^{A_{2j}} \label{eq:A.2.1}
\end{equation}
is the combination of terms of the form 
\begin{equation*}
\frac{1}{(2j)!}\bigl(\mu_Ao^A\bigr)^r\bigl(\mu_B\iota^B\bigr)^k\bigl(\nu_C
o^C\bigr)^l\bigl(\nu_D\iota^D\bigr)^{2j-r-k-l}
\end{equation*}
with non-negative integer coefficients. However, in such a term the number of 
the $o^A$ spinors must be $j+s$, and the number of the $\iota^A$ spinors must 
be $j-s$, and hence $l=j+s-r$. In a similar way, the number of the $\mu_A$ 
spinors is $j-m$ and that of $\nu_A$ is $j+m$, and hence $k=j-m-r$. 
Therefore, for given $r$, the structure of the terms in (\ref{eq:A.2.1}) is 
\begin{equation}
\frac{1}{(2j)!}\bigl(\mu_Ao^A\bigr)^r\bigl(\mu_B\iota^B\bigr)^{j-m-r}\bigl(
\nu_Co^C\bigr)^{j+s-r}\bigl(\nu_D\iota^D\bigr)^{r+m-s}. \label{eq:A.2.2}
\end{equation}
$r$ takes its minimum value when all the $\nu_A$ spinors are contracted 
with $o^A$ (but no $\iota^A$) spinors; and in this case, if $s-m\geq0$, 
the number of the $o^A$ spinors that remain to contract with the $\mu_A$ 
spinors is $s-m$. Hence, $r\geq\max\{0,s-m\}$. The maximum value of $r$ 
is the maximal number of contractions $\mu_Ao^A$, which is $j+s$ if $j-m
\geq j+s$, and it is $j-m$ if $j+s\geq j-m$. Therefore, $r\leq{\rm min}
\{j-m,j+s\}$. Finally, to determine the number of the terms of the form 
(\ref{eq:A.2.2}) in (\ref{eq:A.2.1}), let us rewrite (\ref{eq:A.2.2}) in 
the form 
\begin{eqnarray*}
\frac{1}{(2j)!}\!\!\!\!&{}&\!\!\!\!\Bigl(\mu_{A_1}\cdots\mu_{A_r}
 \nu_{A_{r+1}}\cdots\nu_{A_{j+s}}\Bigr)o^{A_1}\cdots o^{A_{j+s}} \\
\times \!\!\!\!&{}&\!\!\!\!\Bigl(\mu_{B_1}\cdots\mu_{B_{j-m-r}}\nu
 _{B_{j-m-r+1}}\cdots\nu_{B_{j-s}}\Bigr)\iota^{B_1}\cdots\iota^{B_{j-s}}. 
\end{eqnarray*}
The number of ways in which the $\mu_A$ and the $\nu_A$ spinors can be 
chosen in this manner is ${j-m}\choose{r}$ and ${j+m}\choose{r+m-s}$, 
respectively; and each of these choices can be contracted with the $o^A$ 
spinors in $(j+s)!$ ways and with the $\iota^B$ spinors in $(j-s)!$ ways. 
Thus, their total number is ${j-m}\choose{r}$${j+m}\choose{r+m-s}$$(j+s)!
(j-s)!$. Hence 
\begin{eqnarray}
\phi_{s,j,m}=\!\!\!\!&{}\!\!\!\!&\frac{(j+m)!(j-m)!(j+s)!(j-s)!}{(2j)!}
  \nonumber \\
\!\!\!\!&{}\!\!\!\!&\times\sum_r\frac{(\mu_Ao^A)^r(\mu_B\iota^B)^{j-m-r}
  (\nu_Co^C)^{j+s-r}(\nu_D\iota^D)^{r+m-s}}{r!(j-m-r)!(j+s-r)!(r+m-s)!},
  \label{eq:A.2.3a}
\end{eqnarray}
where $\max\{0,s-m\}\leq r\leq\min\{j-m,j+s\}$. In the exceptional case 
$r=j+s$, and hence (\ref{eq:A.2.3a}) reduces to 
\begin{equation}
\phi_{s,j}=\bigl(\mu_Ao^A\bigr)^{j+s}\bigl(\mu_B\iota^B\bigr)^{j-s};
\label{eq:A.2.3b}
\end{equation}
which can be derived directly even from (\ref{eq:A.2.1}), too. 

Next, we determine the explicit form of the contractions $\mu_Ao^A$, 
$\nu_Ao^A$, $\mu_A\iota^A$ and $\nu_A\iota^A$. (These are not only the 
factors in (\ref{eq:A.2.3a}), but also these are the eigenfunctions $\phi
_{s,\frac{1}{2},m}$ with $(s,m)=(1/2,-1/2)$, $(1/2,1/2)$, $(-1/2,-1/2)$ and 
$(-1/2,1/2)$, respectively.) First, suppose that $\gamma_{AB}$ is not null. 
Then 
\begin{equation*}
\gamma_{AB}=\frac{1}{\sqrt{2}}\left(\begin{array}{cc}
                     -\alpha_1+{\rm i}\alpha_2 & \alpha_3-{\rm i}\lambda\\
                       \alpha_3-{\rm i}\lambda & \alpha_1+{\rm i}\alpha_2\\
                  \end{array}\right),
\end{equation*}
and hence the solution of $\sqrt{2}\gamma_{AB}=\mu_A\nu_B+\mu_B\nu_A$ is 
given by 
\begin{eqnarray}
&{}&\mu_A=\mu_1O_A-\mu_0I_A=-\bigl(\xi_{\mp}\exp[{\rm i}\alpha]\,
  O_A+I_A\bigr)\mu_0, \label{eq:A.2.4a} \\
&{}&\nu_A=\nu_1O_A-\nu_0I_A=\frac{1}{2}\sqrt{1-\alpha^2_3}\bigl(
  \xi_{\pm}\,O_A+\exp[-{\rm i}\alpha]\,I_A\bigr)\,\mu_0^{-1}; 
  \label{eq:A.2.4b}
\end{eqnarray}
where $\mu_0:=\mu_AO^A$ is an arbitrary nonzero constant and $\xi_{\pm}$ 
has been defined by (\ref{eq:2.3.3}). However, this solution gives $\nu_A
\mu^A=\mp\sqrt{1-\lambda^2-2{\rm i}\lambda\alpha_3}$, and hence, to be 
compatible with our sign convention in (\ref{eq:2.2.7}), we must choose 
$\xi_+$ in $\mu_1$ (and hence $\xi_-$ in $\nu_1$). Therefore, 
\begin{eqnarray}
&{}&\mu_Ao^A=-{\rm i}\frac{\xi-\xi_+}{\sqrt{1+\xi\bar\xi}}\mu_0
    \exp[{\rm i}\alpha], \hskip 20pt 
    \nu_Ao^A=\frac{\rm i}{2}\sqrt{1-\alpha^2_3}\frac{\xi-\xi_-}
    {\sqrt{1+\xi\bar\xi}}\mu_0^{-1}, \label{eq:A.2.5a}\\
&{}&\mu_A\iota^A=-{\rm i}\xi_+\frac{\bar\xi-\xi_-}{\sqrt{1+\xi\bar\xi}}
     \mu_0, \hskip 20pt
     \nu_A\iota^A=\frac{\rm i}{2}\xi_-\sqrt{1-\alpha^2_3}\frac{\bar\xi-
     \xi_+}{\sqrt{1+\xi\bar\xi}}(\mu_0\exp[{\rm i}\alpha])^{-1}. 
     \label{eq:A.2.5b}
\end{eqnarray}
In the exceptional case, i.e. when $\lambda=1$ and $\alpha_3=0$, the 
components of the principal spinor $\mu_A$ are $\mu_0:=\mu_AO^A=\pm
({\rm i}/\sqrt{2})\exp[-{\rm i}\alpha/2]$ and $\mu_1:=\mu_AI^A=\mp(1/
\sqrt{2})\exp[{\rm i}\alpha/2]$. Hence, the corresponding solutions $\phi
_{1/2,1/2}$ and $\phi_{-1/2,1/2}$, respectively, are 
\begin{equation}
\mu_Ao^A=\pm\frac{1}{\sqrt{2}}\frac{\xi+{\rm i}}{\sqrt{1+
  \xi\bar\xi}}\exp[\frac{\rm i}{2}\alpha], \hskip 20pt 
\mu_A\iota^A=\mp\frac{\rm i}{\sqrt{2}}\frac{\bar\xi+{\rm i}}
  {\sqrt{1+\xi\bar\xi}}\exp[-\frac{\rm i}{2}\alpha]. \label{eq:A.2.5c}
\end{equation}
Therefore, the eigenfunctions $\phi_{s,j,m}$ and $\phi_{s,j}$ are polynomials 
in $\xi$ and $\bar\xi$ (with the overall order $2j$) and divided by 
$(1+\xi\bar\xi)^j$. Thus, in particular, all these are bounded on ${\cal S}$ 
and smooth on ${\cal S}$ minus the `north pole' of the coordinate system 
$(\xi,\bar\xi)$.

\subsubsection{Orthogonality properties and the factors of normalization}
\label{sub-A.2.2}

To find the factor of normalization, and also to clarify whether or not the 
functions $\phi_{s,j,m}$ and $\phi_{s,j}$ form orthogonal systems, we need 
Lemma 4.15.86 of \cite{PR}, viz. that 
\begin{equation}
\int o_{A_1}\cdots o_{A_k}\iota^{B_1}\cdots\iota^{B_k}{\rm d}{\cal S}=
\frac{4\pi}{k+1}\delta^{(B_1}_{A_1}\cdots\delta^{B_k)}_{A_k}; \label{eq:A.2.6}
\end{equation}
and that the integral is zero if the number of $o_A$ and $\iota^B$ spinors 
under the integral is different. Here, and in the rest of this Appendix, 
${\cal S}$ is assumed to be a \emph{unit} sphere. 

Let $\phi_{s,j,m}$ and $\phi_{s',j',m'}$ be two eigenfunctions, where the latter 
is built from $\phi'_{{\uA}_1...{\uA}_{2j'}}$ according to (\ref{eq:2.2.6}) but in 
which $m$ is replaced by $m'$. Recalling that $2\sigma_0^{A'B}\sigma^0_{BB'}
=\delta^{A'}_{B'}$ and that the Hermitean adjoint e.g. of a spinor $\phi_A$ 
is defined by $\phi^\dagger_A:=\bar\phi_{A'}\sqrt{2}\sigma^{A'}_{0\,A}$, we have 
that 
\begin{eqnarray*}
\!\!\!\!&{}\!\!\!\!&\langle\phi_{s',j',m'}\, ,\phi_{s,j,m}\rangle \\
\!\!\!\!&{}\!\!\!\!&=\int\bar\phi'_{D'_1...D'_{2j'}}2\sigma_0^{D'_1D_1}\sigma
  ^0_{D_1A'_1}\cdots2\sigma_0^{D'_{2j'}D_{2j'}}\sigma^0_{D_{2j'}A'_{2j'}}\bar o^{A'_1}
  \cdots\bar o^{A'_{j'+s'}}\bar\iota^{A'_{j'+s'+1}}\cdots\bar\iota^{A'_{2j'}} \\
\!\!\!\!&{}\!\!\!\!&\hskip 20pt \times\phi_{A_1...A_{2j}}o^{A_1}\cdots 
  o^{A_{j+s}}\iota^{A_{j+s+1}}\cdots\iota^{A_{2j}}\,{\rm d}{\cal S} \\
\!\!\!\!&{}\!\!\!\!&=(-)^{j+s}{\phi'}^\dagger_{D_1...D_{j'+s'}}{}
  ^{D_{j'+s'+1}...D_{2j'}}\,\phi^{A_1...A_{j+s}}{}_{A_{j+s+1}...A_{2j}} \\
\!\!\!\!&{}\!\!\!\!&\hskip 20pt \times\int o_{A_1}\cdots o_{A_{j+s}}
 o_{D_{j'+s'+1}}\cdots o_{D_{2j'}}\iota^{D_1}\cdots\iota^{D_{j'+s'}}
 \iota^{A_{j+s+1}}\cdots\iota^{A_{2j}}\,{\rm d}{\cal S}. 
\end{eqnarray*}
However, as we noted above, this integral can be non-zero only if the 
number of the $o_A$ and $\iota^D$ spinors is the same, i.e. if $s'=s$. Hence, 
by (\ref{eq:A.2.6}), this integral is 
\begin{equation*}
(-)^{j+s}\delta_{ss'}\,{\phi'}^\dagger_{D_1...D_{j'+s}}{}^{A_{j+s+1}...A_{j+j'}}\,\phi
^{A_1...A_{j+s}}{}_{D_{j'+s+1}...D_{j+j'}}\frac{4\pi}{j+j'+1}\delta^{(D_1}_{A_1}
\cdots\delta^{D_{j+j'})}_{A_{j+j'}}.
\end{equation*}
But since both $\phi_{A_1...A_{2j}}$ and $\phi'_{A_1...A_{2j'}}$ are totally 
symmetric, a term in this expression is non-zero precisely when every 
index of one spinor is contracted with some index of the other spinor. 
Hence, $j=j'$. Therefore, 
\begin{equation}
\langle\phi_{s',j',m'}\, ,\phi_{s,j,m}\rangle=(-)^{2j}\delta_{ss'}\delta_{jj'}4\pi 
\frac{(j-s)!(j+s)!}{(2j+1)!}\,{\phi'}^\dagger_{A_1...A_{2j}}\phi^{A_1...A_{2j}}.
\label{eq:A.2.7}
\end{equation}
Finally, rewriting ${\phi'}^\dagger_{A_1...A_{2j'}}$ by the complex conjugate 
spinor and recalling that $\sqrt{2}\sigma_0^{AA'}=O^A\bar O^{A'}+I^A\bar I
^{A'}$, we obtain 
\begin{eqnarray}
\langle\phi_{s',j',m'}\, ,\phi_{s,j,m}\rangle\!\!\!\!&=\!\!\!\!&\delta_{ss'}
  \delta_{jj'}4\pi\frac{(j-s)!(j+s)!}{(2j+1)!} \nonumber\\
\!\!\!\!&{}\!\!\!\!&\times\bar\phi'_{A'_1...A'_{2j}}(\bar O^{A'_1}O^{A_1}+\bar I
  ^{A'_1}I^{A_1})\cdots(\bar O^{A'_{2j}}O^{A_{2j}}+\bar I^{A'_{2j}}I^{A_{2j}})\phi
  _{A_1...A_{2j}} \nonumber\\
\!\!\!\!&=\!\!\!\!&\delta_{ss'}\delta_{jj'}4\pi\frac{(j-s)!(j+s)!}
 {(2j+1)!}\sum^{2j}_{k=0}{{2j}\choose{k}}\Phi_{k,j,m}\bar\Phi_{k,j,m'}, 
 \label{eq:A.2.8}
\end{eqnarray}
where $\Phi_{k,j,m}$ was defined by (\ref{eq:2.4.4}). If in $\phi_{A_1
...A_{2j}}$ the spinors $\mu_A$ and $\nu_A$ were $O_A$ and $I_A$, 
respectively, then $\langle\phi_{s',j',m'}\, ,\phi_{s,j,m}\rangle$ would be 
proportional to $\delta_{mm'}$, too, and the whole expression would reduce 
to the expression of the norm of the not normalized spin weighted 
spherical harmonics ${}_sZ(j,m)$ \cite{PR}. However, with our $\mu_A$ 
and $\nu_A$, the eigenfunctions $\phi_{s,j,m}$ and $\phi_{s,j,m'}$, $m\not=
m'$, are \emph{not} orthogonal to each other. Nevertheless, (\ref{eq:A.2.8}) 
provides the factor of normalization $N_{s,j,m}$: 
\begin{equation}
N^{-2}_{s,j,m}:=\langle\phi_{s,j,m},\phi_{s,j,m}\rangle=4\pi\frac{(j-s)!(j+s)!}
{(2j+1)!}\sum^{2j}_{k=0}{{2j}\choose{k}}\vert\Phi_{k,j,m}\vert^2.\label{eq:A.2.9}
\end{equation}
To see its dependence on the parameters $\alpha_3$ and $\lambda$, we should 
find the explicit form of $\Phi_{k,j,m}$. We determine this in the next 
subsection. 

The previous analysis is valid even in the exceptional case. The only 
difference between the generic and exceptional cases is that now $m=m'=-j$, 
and hence, according to (\ref{eq:A.2.8}), the eigenfunctions $\phi_{s,j}$ 
form an orthogonal system. Since, in the exceptional case, the modulus of 
the spinor components, $\mu_0:=\mu_AO^A$ and $\mu_1:=\mu_AI^A$, 
is $1/\sqrt{2}$, the factor of normalization $N_{s,j}=N_{s,j,-j}$, given by 
(\ref{eq:A.2.9}), can be written as 
\begin{equation}
N^{-2}_{s,j}=4\pi\frac{(j-s)!(j+s)!}{(2j+1)!}\sum_{k=0}^{2j}{{2j}\choose{k}}
\vert\mu_0\vert^{2k}\,\vert\mu_1\vert^{2(2j-k)}=4\pi\frac{(j-s)!(j+s)!}
{(2j+1)!}, \label{eq:A.2.10}
\end{equation}
which is a pure number, independently of the (only) free parameter $\alpha$ 
in $\mu_0$ and $\mu_1$.

\subsubsection{The continuity of the standard deviations}
\label{sub-A.2.3}

(\ref{eq:A.2.3a}) gives the structure of $\Phi_{k,j,m}$, too, if $o^A$ is 
replaced by $O^A$, $\iota^A$ by $I^A$, and $s$ by $k-j$. Thus, using 
(\ref{eq:A.2.4a})-(\ref{eq:A.2.4b}) and taking into account $\xi_+\xi_-=-1$ 
(see (\ref{eq:2.3.3})), we find that 
\begin{eqnarray}
\Phi_{k,j,m}\!\!\!\!&=\!\!\!\!&(-)^k\mu_0^{-2m}\bigl(\frac{1}{2}\sqrt{1-
 \alpha^2_3}\bigr)^{j+m}\bigl(\exp[{\rm i}\alpha]\bigr)^{j-m-k}(\xi_-)^{2m-k} 
  \nonumber \\
\!\!\!\!&{}\!\!\!\!&\times\frac{k!(2j-k)!}{(2j)!}\sum_r(-)^r{{j-m}\choose{r}}
 {{j+m}\choose{k-r}}(\xi_-)^{2r}, \label{eq:A.2.11} 
\end{eqnarray}
where ${\rm max}\{0,k-j-m\}\leq r\leq{\rm min}\{j-m,k\}$. However, by 
equations (\ref{eq:A.2.9}), (\ref{eq:2.4.3b}) and (\ref{eq:2.4.5}), in the 
expression of $\langle{\bf J}(\beta)\rangle_\phi$ and $\langle({\bf J}(\beta))
^2\rangle_\phi$ it is only the absolute value of 
\begin{equation}
S_{k,j,m}:=(\xi_-)^{-k}\sum_r(-)^r{{j-m}\choose{r}}{{j+m}\choose{k-r}}(\xi_-)
^{2r} \label{eq:A.2.12}
\end{equation}
that appears. In fact, in terms of this, $\langle W_{s,j,m},{\bf J}_3W_{s,j,m}
\rangle$ and $\langle{\bf J}_3W_{s,j,m},{\bf J}_3W_{s,j,m}\rangle$ take the 
form 
\begin{eqnarray}
\langle W_{s,j,m},{\bf J}_3W_{s,j,m}\rangle\!\!\!\!&=\!\!\!\!&-\hbar
  \frac{\sum_{k=0}^{2j}(j-k)k!(2j-k)!\vert S_{k,j,m}\vert^2}{\sum_{k=0}^{2j}k!
  (2j-k)!\vert S_{k,j,m}\vert^2}, \label{eq:A.2.13a} \\
\langle{\bf J}_3W_{s,j,m},{\bf J}_3W_{s,j,m}\rangle\!\!\!\!&=\!\!\!\!&\hbar^2
  \frac{\sum_{k=0}^{2j}(j-k)^2k!(2j-k)!\vert S_{k,j,m}\vert^2}{\sum_{k=0}^{2j}k!
  (2j-k)!\vert S_{k,j,m}\vert^2}. \label{eq:A.2.13b}
\end{eqnarray}
These depend on $\alpha_3$ and $\lambda$ only through $\xi_-$ according to 
(\ref{eq:2.3.3}) via $S_{k,j,m}$, but they do not depend on $\alpha$ or 
$\mu_0$ (as they should not). 

To see the structure of $S_{k,j,m}$ and its dependence on $\alpha_3$ and 
$\lambda$, let us recall that the range of the summation in (\ref{eq:A.2.12}) 
is $\max\{0,k-j-m\}\leq r\leq\min\{j-m,k\}$. Hence, we should split the range 
of $k$ in (\ref{eq:A.2.13a})-(\ref{eq:A.2.13b}) into three disjoint domains: 
i. when $0\leq k< j-\vert m\vert$, then $r=0,...,k$; ii.a. when 
$j-\vert m\vert\leq k\leq j+\vert m\vert$ and $m\geq0$, then $r=0,...,j-m$, 
ii.b. if $j-\vert m\vert\leq k\leq j+\vert m\vert$ and $m<0$, then 
$r=k-j-m,...,k$; iii. when $j+\vert m\vert< k\leq 2j$, then $r=
k-j-m,...,j-m$. Hence, redefining $r$ and its range if needed, in the 
respective cases 
\begin{eqnarray}
{\rm i.}  &{}&S_{k,j,m}=(\xi_-)^{-k}\sum_{r=0}^k(-)^r{{j-m}\choose{r}}{{j+m}
  \choose{k-r}}(\xi_-)^{2r},  \label{eq:A.2.14a} \\
{\rm ii.a.} &{}&S_{k,j,m}=(\xi_-)^{-k}\sum_{r=0}^{j-m}(-)^r{{j-m}\choose{r}}{{j+m}
  \choose{k-r}}(\xi_-)^{2r}, \label{eq:A.2.14b} \\
{\rm ii.b.} &{}&S_{k,j,m}=(-)^{k-j-m}(\xi_-)^{k-2(j+m)}\sum_{r=0}^{j+m}(-)^r
  {{j-m}\choose{2j-k-r}}{{j+m}\choose{r}}(\xi_-)^{2r}, \label{eq:A.2.14c} \\ 
{\rm iii.} &{}&S_{k,j,m}=(-)^{k-j-m}(\xi_-)^{k-2(j+m)}\sum_{r=0}^{2j-k}(-)^r
  {{j-m}\choose{2j-k-r}}{{j+m}\choose{r}}(\xi_-)^{2r}. \label{eq:A.2.14d}
\end{eqnarray}
Next we show that these expressions admit a rather non-trivial, hidden 
symmetry: for given $k,j,m$ let us define $\tilde k:=2j-k$, and clearly $k=
0,...,j-\vert m\vert-1$ precisely when $\tilde k=j+\vert m\vert+1,...,2j$; 
and $k=j-\vert m\vert,...,j+\vert m\vert$ precisely when $\tilde k=j-\vert m
\vert,...,j+\vert m\vert$. Then e.g. for $k=0,...,j-\vert m\vert-1$ and $m
\geq0$ by (\ref{eq:A.2.14a}) and (\ref{eq:A.2.14d}) with the notation 
$\bar r:=r-k+j-m$ we have that 
\begin{eqnarray}
S_{k,j,-m}\!\!\!\!&=\!\!\!\!&(\xi_-)^{-k}\sum_{r=0}^k(-)^r{{j+m}\choose{r}}
  {{j-m}\choose{k-r}}(\xi_-)^{2r} \nonumber\\
\!\!\!\!&=\!\!\!\!&(\xi_-)^{\tilde k-2j}\sum_{\bar r=\tilde k-j-m}^{j-m}(-)
  ^{\bar r+j+m-\tilde k}{{j+m}\choose{\bar r-\tilde k+j+m}}{{j-m}\choose{j-m-
  \bar r}}(\xi_-)^{2(\bar r-\tilde k+j+m)} \nonumber\\
\!\!\!\!&=\!\!\!\!&(-)^{j+m-\tilde k}(\xi_-)^{2m-\tilde k}\sum_{\bar r=\tilde k-j-m}
  ^{j-m}(-)^{\bar r}{{j+m}\choose{\tilde k-\bar r}}{{j-m}\choose{\bar r}}
  (\xi_-)^{2\bar r} \nonumber\\
\!\!\!\!&=\!\!\!\!&(-)^{j+m-\tilde k}(\xi_-)^{2m}S_{\tilde k,j,m}. \label{eq:A.2.15}
\end{eqnarray}
Analogous calculations in the other cases show that $S_{k,j,-m}=(-)^{j+m-\tilde k}
(\xi_-)^{2m}S_{\tilde k,j,m}$ holds in general. 

This symmetry of $S_{k,j,m}$ yields that the common denominator in 
(\ref{eq:A.2.13a}) and (\ref{eq:A.2.13b}) is 
\begin{equation*}
\sum_{k=0}^{2j}k!(2j-k)!\vert S_{k,j,-m}\vert^2=\vert\xi_-\vert^{4m}\Bigl(
\sum_{k=0}^{2j}k!(2j-k)!\vert S_{k,j,m}\vert^2\Bigr), \label{eq:A.2.16}
\end{equation*}
while their numerators, respectively, are 
\begin{eqnarray*}
&{}&\sum_{k=0}^{2j}(j-k)k!(2j-k)!\vert S_{k,j,-m}\vert^2=-\vert\xi_-\vert^{4m}
  \Bigl(\sum_{k=0}^{2j}(j-k)k!(2j-k)!\vert S_{k,j,m}\vert^2\Bigr), 
  \label{eq:A.2.17a}\\
&{}&\sum_{k=0}^{2j}(j-k)^2k!(2j-k)!\vert S_{k,j,-m}\vert^2=\vert\xi_-\vert
  ^{4m}\Bigl(\sum_{k=0}^{2j}(j-k)^2k!(2j-k)!\vert S_{k,j,m}\vert^2\Bigr). 
  \label{eq:A.2.17b}
\end{eqnarray*}
These immediately imply that $\langle W_{s,j,-m},{\bf J}_3 W_{s,j,-m}\rangle=-
\langle W_{s,j,m},{\bf J}_3 W_{s,j,m}\rangle$, as we expected, and $\langle
{\bf J}_3W_{s,j,-m},{\bf J}_3 W_{s,j,-m}\rangle=\langle{\bf J}_3W_{s,j,m},{\bf J}
_3W_{s,j,m}\rangle$. 

Next we show that $\langle{\bf J}_3W_{s,j,m},{\bf J}_3 W_{s,j,m}\rangle$ is a 
continuous function of $\alpha_3$, although the expectation value $\langle 
W_{s,j,m},{\bf J}_3W_{s,j,m}\rangle$ has a jump at $\alpha_3=0$ for $\lambda>1$. 

$\xi_-$, as a function of $\alpha_3$ (see (\ref{eq:2.3.3})), is \emph{not} 
continuous at $\alpha_3=0$ for $\lambda>1$: its $\alpha_3\to0$ limit from 
the left is $-{\rm i}(\lambda+\sqrt{\lambda^2-1})$, while from the right it 
is $-{\rm i}(\lambda-\sqrt{\lambda^2-1})$. Hence, $S_{k,j,m}$ is not continuous 
there. Now we calculate the left limit of $S_{k,j,m}$, denoted by $S_{k,j,m}^-$, 
and relate it to the right limit $S_{k,j,m}^+$ of $S_{k,j,m}$. The key 
observation is that $\lambda+\sqrt{\lambda^2-1}=(\lambda-\sqrt{\lambda^2-1})
^{-1}$, but the strategy of the calculation is the same that we followed in the 
derivation of (\ref{eq:A.2.15}). In particular, for $k=0,...,j-m-1$, $m\geq0$, 
with the notation $\bar r:=j-m-r$ the main steps are 
\begin{eqnarray}
S^-_{k,j,m}\!\!\!\!&=\!\!\!\!&(-{\rm i})^{-k}\bigl(\lambda+\sqrt{\lambda^2-1})
  ^{-k}\sum_{r=0}^k(-)^r{{j-m}\choose{r}}{{j+m}\choose{k-r}}\bigl(-{\rm i}
  (\lambda+\sqrt{\lambda^2-1})\bigr)^{2r} \nonumber \\
\!\!\!\!&=\!\!\!\!&(-{\rm i})^{\tilde k-2m}\bigl(\lambda-\sqrt{\lambda^2-1})^{-
  \tilde k+2m}\sum_{\bar r=j-m}^{\tilde k-j-m}(-)^{\bar r}{{j-m}\choose{\bar r}}
  {{j+m}\choose{\tilde k-\bar r}}\bigl(-{\rm i}(\lambda-\sqrt{\lambda^2-1})
  \bigr)^{2\bar r} \nonumber \\
\!\!\!\!&=\!\!\!\!&(-)^{j-k}\bigl(\lambda-\sqrt{\lambda^2-1}\bigr)^{2m}S^+
  _{\tilde k,j,m}. \label{eq:A.2.18}
\end{eqnarray}
One can check that $S^-_{k,j,m}=(-)^{j-k}(\lambda-\sqrt{\lambda^2-1})^{2m}S^+
_{\tilde k,j,m}$ holds in the other cases, too. Hence the $\alpha_3\to0$ limit 
from the left of both the denominator and the numerator in (\ref{eq:A.2.13b}) 
is $(\lambda-\sqrt{\lambda^2-1})^{4m}$ times their $\alpha_3\to0$ limit from 
the right, and hence the left and right limits of $\langle{\bf J}_3W_{s,j,m},
{\bf J}_3W_{s,j,m}\rangle$ coincide. Similar argumentation confirms that 
$\langle W_{s,j,m},{\bf J}_3W_{s,j,m}\rangle$ changes sign at $\alpha_3=0$ for 
$\lambda>1$, i.e. it jumps there. 

Since $\langle{\bf J}_3W_{s,j,m},{\bf J}_3W_{s,j,m}\rangle$ and $(\langle W
_{s,j,m},{\bf J}_3W_{s,j,m}\rangle)^2$ are continuous and $\Delta_\phi{\bf J}
(\alpha)=\lambda\Delta_\phi{\bf J}(\beta)$ holds, (\ref{eq:2.4.6}) implies 
that the standard deviations, both for ${\bf J}(\alpha)$ and ${\bf J}
(\beta)$, are continuous functions of the parameters $\alpha_3$ and 
$\lambda$.

\subsubsection{The $\lambda\to0$ and $\lambda\to\infty$ limits of the 
standard deviations}
\label{sub-A.2.4}

First consider the $\lambda\to0$ limit. By (\ref{eq:2.3.3}), in this limit, 
\begin{equation*}
\xi_-=-\sqrt{\frac{1-\alpha_3}{1+\alpha_3}}+O\bigl(\lambda^2\bigr)+{\rm i}
O\bigl(\lambda\bigr). 
\end{equation*}
Hence, by (\ref{eq:A.2.12}), all $S_{k,j,m}$ are bounded in this limit, 
implying that $\langle{\bf J}_3W_{s,j,m},{\bf J}_3W_{s,j,m}\rangle$ $=\langle
{\bf J}_3W_{s,j,m},{\bf J}_3W_{s,j,m}\rangle\vert_{\lambda=0}+O(\lambda)$. Since, 
as we saw in subsection \ref{sub-2.2}, the expectation value is also finite 
in this limit, the standard deviation for ${\bf J}(\beta)$ is finite. Then 
$\Delta_\phi{\bf J}(\alpha)=\lambda\Delta_\phi{\bf J}(\beta)$ implies 
that both the standard deviation for ${\bf J}(\alpha)$ and the product 
uncertainty, $\Delta_\phi{\bf J}(\alpha)\Delta_\phi{\bf J}(\beta)$, tend to 
zero as $\lambda$ in the $\lambda\to0$ limit. 

In the $\lambda\to\infty$ limit, (\ref{eq:2.3.3}) gives that 
\begin{eqnarray}
&{}&\xi_-=-\frac{1}{2}\sqrt{1-\alpha^2_3}\bigl({\rm i}\lambda^{-1}+\alpha_3
  \lambda^{-2}+O(\lambda^{-3})\bigr), \hskip 20pt {\rm if}\,\,\alpha_3>0, 
    \label{eq:A.2.19a} \\
&{}&\xi_-=-\frac{2}{\sqrt{1-\alpha^2_3}}\bigl({\rm i}\lambda+\vert\alpha_3
  \vert+O(\lambda^{-1})\bigr). \hskip 50pt {\rm if}\,\,\alpha_3\leq0. 
    \label{eq:A.2.19b}
\end{eqnarray}
To calculate the standard deviations in this limit, let us recall from the 
previous subsection that $\langle W_{s,j,-m},{\bf J}_3 W_{s,j,-m}\rangle=
-\langle W_{s,j,m},{\bf J}_3 W_{s,j,m}\rangle$ and $\langle{\bf J}_3W_{s,j,-m},
{\bf J}_3 W_{s,j,-m}\rangle=\langle{\bf J}_3W_{s,j,m},{\bf J}_3W_{s,j,m}\rangle$, 
and hence the standard deviation of ${\bf J}_3$ in the states $W_{s,j,m}$ and 
$W_{s,j,-m}$ is the same. Therefore, it is enough to consider the $m=0$ and 
$m>0$ disjoint cases. 

First suppose that $m=0$. Then, with the notations of the previous 
subsection, the hidden symmetry (\ref{eq:A.2.15}) yields that $S_{k,j,0}=
(-)^{k-j}S_{\tilde k,j,0}$. Hence the denominator and the numerator, respectively, 
in (\ref{eq:A.2.13b}) are 
\begin{equation*}
(j!)^2\vert S_{j,j,0}\vert^2+\sum_{k=0}^{j-1}k!(2j-k)!\vert S_{k,j,0}
\vert^2, \hskip 30pt 
2\sum_{k=0}^{j-1}(j-k)^2k!(2j-k)!\vert S_{k,j,0}\vert^2;
\end{equation*}
while the numerator in (\ref{eq:A.2.13a}) is vanishing. Therefore, the square 
of the standard deviation of ${\bf J}_3$ in the states $W_{s,j,0}$ is simply 
the quotient of these two. Thus the only thing to do is to determine the 
leading order terms of these two expressions in the cases (\ref{eq:A.2.19a}) 
and (\ref{eq:A.2.19b}). However, in both cases 
\begin{equation*}
S_{k,j,0}=(\xi_-)^{-k}\sum_{k=0}^k(-)^r{{j}\choose{r}}{{j}\choose{k-r}}
(\xi_-)^{2r}=\bigl(\frac{2{\rm i}}{\sqrt{1-\alpha^2_3}}\bigr)^k
{{j}\choose{k}}\lambda^k+O(\lambda^{k-1}),
\end{equation*}
yielding that the order of the leading term of the numerator is $k=j-1$, 
and in the denominator it is $k=j$. Therefore, 
\begin{equation*}
\langle{\bf J}_3W_{s,j,0},{\bf J}_3 W_{s,j,0}\rangle=\frac{1}{2}j(j+1)\bigl(
1-\alpha^2_3\bigr)\lambda^{-2}+O(\lambda^{-3});
\end{equation*}
and hence, by $m=0$, the standard deviation for ${\bf J}(\beta)$ tends to 
zero as $1/\lambda$, and that for ${\bf J}(\alpha)$ remains finite. Thus, 
by (\ref{eq:2.4.6}) in the states $\phi$ with eigenvalue $m=0$, the product 
uncertainty $\Delta_\phi{\bf J}(\alpha)\,\Delta_\phi{\bf J}(\beta)=\lambda
(\Delta_\phi{\bf J}(\beta))^2$ tends to zero as $1/\lambda$ in the $\lambda
\to\infty$ limit. 

Next suppose that $m>0$. If $j$ is an integer, then the common denominator in 
(\ref{eq:A.2.13a}) and (\ref{eq:A.2.13b}) is 
\begin{eqnarray*}
D:=\!\!\!\!&{}\!\!\!\!&\sum_{k=0}^{j-m-1}k!(2j-k)!\vert S_{k,j,m}\vert^2+
  \sum_{k=j-m}^{j-1}k!(2j-k)!\vert S_{k,j,m}\vert^2\\
\!\!\!\!&{}\!\!\!\!&+(j!)^2\vert S_{j,j,m}\vert^2+\sum_{k=j+1}^{j+m}k!(2j-k)!
  \vert S_{k,j,m}\vert^2+\sum_{k=j+m+1}^{2j}k!(2j-k)!\vert S_{k,j,m}\vert^2.
\end{eqnarray*}
Then, using (\ref{eq:A.2.14a}), (\ref{eq:A.2.14b}), (\ref{eq:A.2.14d}) and 
(\ref{eq:A.2.19a}), it is a tedious but routine calculation to determine the 
leading order terms in $D$ when $\alpha_3>0$. It is 
\begin{equation*}
D=(j-m)!(j+m)!\bigl(\frac{2}{1-\alpha_3^2}\bigr)^{j+m}\lambda^{2(j+m)}+
O(\lambda^{2(j+m-1)}).
\end{equation*}
We obtain the same expression when $j$ is a half-odd-integer. Similar 
calculations show that the numerators $N_1$ and $N_2$ in (\ref{eq:A.2.13a}) 
and (\ref{eq:A.2.13b}), respectively, are 
\begin{eqnarray*}
&{}&N_1=-m(j+m)!(j-m)!\bigl(\frac{2}{1-\alpha_3^2}\bigr)^{j+m}\lambda
 ^{2(j+m)}+O(\lambda^{2(j+m-1)}), \\
&{}&N_2=m^2(j+m)!(j-m)!\bigl(\frac{2}{1-\alpha_3^2}\bigr)^{j+m}\lambda
 ^{2(j+m)}+O(\lambda^{2(j+m-1)}).
\end{eqnarray*}
If $\alpha_3\leq0$, then, using (\ref{eq:A.2.19b}), analogous calculations 
yield the same expressions for $D$, $N_1$ and $N_2$ except that $m$ should be 
replaced by $-m$. Hence, 
\begin{equation*}
\langle W_{s,j,m},{\bf J}_3W_{s,j,m}\rangle={\rm sign}(\alpha_3)\,m\hbar
+O(\lambda^{-2}), \hskip 20pt
\langle{\bf J}_3W_{s,j,m},{\bf J}_3W_{s,j,m}\rangle=m^2\hbar^2+O(\lambda^{-2});
\end{equation*}
which, by (\ref{eq:2.4.6}), imply that, in the states $\phi$ with eigenvalue 
$m>0$, the standard deviation for ${\bf J}(\beta)$ tends to zero as $1/
\lambda$, and that for ${\bf J}(\alpha)$ remains finite. These imply that 
the product uncertainty $\Delta_\phi{\bf J}(\alpha)\,\Delta_\phi{\bf J}
(\beta)=\lambda(\Delta_\phi{\bf J}(\beta))^2$ tends to zero as $1/\lambda$ 
in the $\lambda\to\infty$ limit. 


\end{document}